\documentclass[10pt, aps, prd, onecolumn, superscriptaddress, nofootinbib, notitlepage, floatfix]{revtex4-1}
\usepackage{latexsym,amsmath,amssymb,amstext}
\usepackage{slashed}
\usepackage{epsfig}
\usepackage{float,xcolor}
\usepackage{graphicx}
\usepackage{dcolumn}
\usepackage{comment}

\newcommand{\be}{\begin{equation}}
\newcommand{\ee}{\end{equation}}
\newcommand{\bea}{\begin{eqnarray}}
\newcommand{\eea}{\end{eqnarray}}
\newcommand\bef{\begin{figure*}}
\newcommand\eef[1]{\label{fg:#1}\end{figure*}}
\newcommand\beq{\begin{equation}}
\newcommand\eeq[1]{\label{#1}\end{equation}}
\newcommand\beqa{\begin{eqnarray}}
\newcommand\eeqa[1]{\label{#1}\end{eqnarray}}
\newcommand\bet{\begin{table}}
\newcommand\eet[1]{\label{tb:#1}\end{table}}

\newcommand\fgn[1]{Figure \ref{fg:#1}}
\newcommand\eqn[1]{Eq.\ (\ref{#1})}
\newcommand\scn[1]{Section \ref{sec:#1}}
\newcommand\apx[1]{Appendix \ref{sec:#1}}

\newcommand{\Feyn}[1]{#1\!\!\!\!\slash}
\newcommand{\Pf}{{\rm Pf}}
\newcommand{\Tr}{{\rm Tr }}

\newcommand{\Nu}{\mathcal{V}}
\begin{document}

\date{\today}

\title{Lattice study of two-dimensional SU(2) gauge theories with a single massless Majorana fermion}

\author{Rajamani\ \surname{Narayanan}}
\email{rajamani.narayanan@fiu.edu}
\author{Ray\ \surname{Romero}}
\email{rrome071@fiu.edu}
\affiliation{Department of Physics, Florida International University, Miami, FL 33199}

\begin{abstract}
Massless overlap fermions in the real representation of two dimensional $SU(N_c)$ gauge theories exhibit a mod($2$) index due to the rigidity of its spectrum when viewed as a function of the background gauge field -- lattice gauge fields on a periodic torus come under two classes; ones that have one set of chirally paired zero modes and ones that do not. Focusing on $SU(2)$ and a single Majorana fermion in an integer representation, $J$; we present numerical evidence that shows only one of these classes survives the continuum limit and this depends on the boundary conditions of the fermion and the gauge field. As such, two of the four possible partition functions are zero in the continuum limit. By defining modified partition functions which do not include the zero modes of the overlap fermions in the fermion determinant, we are able to define an expectation value for a fermion bilinear as ratios of two mixed partition functions. This observable is referred to as the topological condensate and has a non-zero expectation value on any finite physical torus and also has a non-zero limit as the size of the torus is taken to infinity. We study the spectral density of fermions and the scaling of the lowest eigenvalue with the size of the torus to show the absence of any spontaneous symmetry breaking but the emergence of zero modes in the infinite volume limit where it is prohibited in finite volume. These results remain the same for $J=1,2,3,4$. These results motivate us to propose an independent plaquette model which reproduces the correct physics in the infinite volume limit using a single partition function.
\end{abstract}

\maketitle

\section{Introduction and summary}\label{sec:intro}

Two dimensional $SU(N_c)$ gauge theories with $N_f$ flavors of Dirac fermions in the fundamental representation serve as interesting examples of conformal field theories (CFTs)~\cite{Delmastro:2021otj}. The $U(N_f)$ flavor symmetry results in the conjecture of $U(N_f)$ WZW model with level $\frac{N_c}{2}$ as the appropriate CFT in the infra-red. This was numerically verified for a sampling of $N_c$ and $N_f$ in~\cite{Karthik:2023gud}.
A continuous flavor symmetry is absent  when there is a single flavor of massless Majorana fermion in a real representation of the gauge group but we still have a discrete $Z_2$ chiral symmetry. Some of these theories are expected to have a massless sector described by certain CFTs in the infra-red~\cite{Delmastro:2021otj}. Let the integer $J$ label the real representation of $SU(2)$ where the dimension of the fermion representation is $(2J+1)$. The theory with $J=1$ is the massless adjoint $SU(2)$ gauge theory and has a mass gap. The particle spectrum has been studied using  Discrete Light Cone Quantization (DLCQ)~\cite{Dempsey:2022uie,Narayanan:2023jmi}. A particular lattice Hamiltonian~\cite{Dempsey:2023fvm} has been used to compute a fermion bilinear condensate. An attempt to compute a fermion bilinear condensate using massive Wilson fermions can be found in~\cite{Bergner:2024ttq}. Arguments for a gap in $J=1,2$ theories and the absence of a gap for $J>3$ are provided in~\cite{Delmastro:2021otj} and numerically verified using DLCQ in~\cite{Narayanan:2023jmi}. 

Majorana-Weyl fermions form the fundamental building block in $SU(N_c)$ gauge theories with fermions in the real representation. A mod($2$) index was introduced in~\cite{Huet:1996pw} and overlap fermions were shown to exhibit this index when it is present. The effect of the boundary conditions of fermions and gauge fields on a periodic torus on the presence and absence of this mod($2$) index was discussed in~\cite{Cherman:2019hbq}. The basic argument for the mod($2$) index introduced in~\cite{Huet:1996pw} is as follows. Let the Euclidean massless Dirac operator in a real representation of the gauge field be
\be
\Feyn{D} = \sum_{j=1}^2 \sigma_j(\partial_j+iA_j);\qquad A_j=A^\dagger_j=-A^*_j;\qquad
\sigma_1 = \begin{pmatrix} 0 & 1 \cr 1 & 0 \end{pmatrix};\quad
\sigma_2 = \begin{pmatrix} 0 & i \cr -i & 0 \end{pmatrix};\quad
\sigma_3 = \begin{pmatrix} -1 & 0 \cr 0 & 1 \end{pmatrix}.
\ee
Then
\be
\{\sigma_3,\Feyn{D}\}=0;\qquad \sigma_1 {\Feyn{D}}^* \sigma_1 = \Feyn{D}
\ee
implies that every eigenvalue is doubly degenerate including zero modes and that every non-zero eigenvalue has a chiral pair. Therefore the number of zero modes mod($4$) is robust under perturbations. A trivial example occurs when we set $A_j=0$ and fermions obey periodic boundary conditions on the torus. Focusing on $SU(2)$, we have $2(2J+1)$ zero momentum modes and small perturbations of the gauge field will lift $4J$ zero modes leaving two zero modes that cannot be lifted under arbitrary {\sl smooth} deformations of the gauge field. If the fermions obey anti-periodic boundary conditions, the mod($2$) index will be trivial and remain so under arbitrary {\sl smooth} deformations of the gauge field. The situation is not so trivial with the gauge fields obey the 't Hooft twisted boundary conditions~\cite{tHooft:1979rtg,tHooft:1980kjq} and this case has been analyzed in~\cite{Cherman:2019hbq}. The outcome in this case is reveresed -- fermions with anti-periodic boundary conditions have a non-trivial mod($2$) index whereas fermions with periodic boundary conditions do not. This argument does not imply all gauge fields obey these properties -- we may have disconnected sectors of gauge fields.

Let $H$ be the Hamiltonian for $SU(2)$ gauge fields with a single massless Majorana fermion and let us summarize the analysis of the spectrum found in~\cite{Cherman:2019hbq} and in~\cite{Dempsey:2023fvm}. We first note that the fermion number operator, $F$ ($F^2=1$), commutes with $H$. The 't Hooft twisted boundary condition can be introduced using the insertion of a center symmetry operator $U$ ($U^2=1$) that commutes with $H$ and $F$. The Hamiltonian along with the center symmetry operator, $U$, and the fermion number operator $F$ form a commuting set.  We will label the simultaneous eigenstates of $H$, $U$ and $F$ by $|E,z,f\rangle$ where $E$ is the eigenvalue of $H$, $z=\pm 1$ and $f=\pm 1$ are the eigenvalues of $U$ and $F$ respectively. The axial symmetry operator, $\Nu$, commutes with $H$ and anti-commutes with $F$ and $U$.  The fermion mass operator, $M$, commutes with $U$ and $F$ but anti-commutes with $\Nu$. We consider four different partition functions
\be
Z_{+-}=\Tr e^{-\frac{H}{T} };\qquad Z_{++} = \Tr \left[F e^{-\frac{H}{T} }\right];\qquad Z_{--} = \Tr \left[Ue^{-\frac{H}{T} }\right];\qquad Z_{-+} = \Tr \left[ UF e^{-\frac{H}{T} }\right].\label{diffpart}
\ee
Noting that
\be
\Nu|E,z,f\rangle = |E,-z,-f\rangle
\ee
is a different state\footnote{This lends itself to a multiverse interpretation~\cite{Cherman:2019hbq,Komargodski_2021,Dempsey:2023fvm} but we will not use it for the discussion in this paper.}, we can conclude that
\be
Z_{++}=Z_{--}=0;\qquad \Tr \left[ M e^{-\frac{H}{T} }\right] = \Tr \left[M UF e^{-\frac{H}{T} }\right]=0.\label{zerocond}
\ee
The fermion bilinear condensate in the limit of $T\to 0$ is defined as the expectation value of the mass operator in the ground state ($E=E_0$) at a fixed $z$ in~\cite{Dempsey:2023fvm}. If the condensate is due to the presence of exact zero modes,
it cannot be calculated as an expectation value with respect to a single partition function defined above but we can consider\footnote{Mixed partition functions were also considered in the context of four dimensional SYM~\cite{Anber:2022qsz,Anber:2024mco,Anber:2025vjo} to enable a computation of the gluino condensate by mixing twisted sectors.}
\be
Z_1=Z_{+-}+Z_{++};\qquad Z_2=Z_{--}+Z_{-+};\qquad Z_3=Z_{+-}+Z_{--};\qquad Z_4=Z_{++}+Z_{-+}.
\ee
Then
\be
\langle M_1\rangle = \frac{\Tr \left[MF e^{-\frac{H}{T} }\right]}{Z_{+-}};\qquad
\langle M_2\rangle = \frac{\Tr \left[MU e^{-\frac{H}{T} }\right]}{Z_{-+}};\qquad
\langle M_3\rangle = \frac{\Tr \left[MU e^{-\frac{H}{T} }\right]}{Z_{+-}};\qquad
\langle M_4\rangle = \frac{\Tr \left[MF e^{-\frac{H}{T} }\right]}{Z_{-+}}.\label{topcond}
\ee
can be non-zero and different at finite temperature or as $T \to 0$. 
The above argument translates to partition functions on an $\ell^2$ torus where $\ell$ is the finite extent of the torus and we use the Lagrangian formalism to study the theory.   We will refer to the density associated with the quantities in \eqn{topcond} as  topological bilinear condensates since they are present on a finite torus due to the mod($2$) index and the mass operator is the {\sl 't Hooft vertex} that absorbs the zero modes. In other words, the $Z_2$ chiral symmetry is explicitly broken by the presence of the zero modes. We will refer such a condensate as a topological condensate. This is different from a fermion bilinear condensate arising from spontaneous symmetry breaking. Consider
\be
\lim_{m\to 0}\lim_{T\to 0} \frac{d}{dm} \ln \left[\Tr\left(e^{-\frac{1}{T}(H  + mM)}\right)\right];\quad{\rm or}
\lim_{m\to 0}\lim_{T\to 0} \frac{d}{dm} \ln \left[\Tr\left(UFe^{-\frac{1}{T} (H + mM)}\right)\right].\label{sponcond}
\ee
If the above expressions result in a non-zero value, we would say that the $Z_2$ chiral symmetry is spontaneously broken~\cite{Banks:1979yr}.

Our aim in this paper is to address the difference between the topological and the spontaneous bilinear condensates. We will perform a numerical study using an Euclidean lattice action and the single massless Majorana fermion will be realized using the overlap formalism. The fermions can either obey periodic or anti-periodic boundary conditions. We will work on an $L^2$ lattice and the $SU(2)$ group valued gauge fields on the links will obey periodic boundary conditions. The twisted boundary conditions will be implemented by switching the sign of the lattice gauge coupling, $\beta$, on one of the $L^2$ plaquettes. The continuum limit will be taken by setting the lattice gauge coupling
\be
\beta = \frac{L^2}{\ell^2};\qquad L\to\infty
\ee
and $\ell$ is the dimensionless extent of the symmetric continuum torus measured in units of the gauge coupling $g$ by $\ell = \ell_p g$. Our paper is organized as follows. 

Given the Hermitian Wilson-Dirac operator, $H_w$, let $R$ be the gauge-field dependent unitary matrix that diagonalizes $H_w$. The unitary operator
\be
V = \sigma_3 R \sigma_3 R^\dagger\label{Vop}
\ee
that compares $\sigma_3$ with a {\sl rotated} $\sigma_3$ by $R$ will play a significant role in the overlap formalism. 
Because the fermions are coupled to a real representation of the gauge group, we will be able to show in \scn{specHw} that the spectrum of $H_w$ comes in pairs, $\pm\lambda_i$. Note that this is not due to the chiral symmetry which is broken by Wilson fermions. 
We will show in \scn{specV} that the spectral representation of $V$
will be of the form 
\be
V\chi_{1j} = e^{i\phi_j}\chi_{1j};\qquad
V\chi_{2j} = e^{-i\phi_j}\chi_{2j};\qquad
V\chi_{3j} = e^{i\phi_j}\chi_{3j};\qquad
V\chi_{4j} = e^{-i\phi_j}\chi_{4j};\qquad 0 < \phi_j < \pi
\ee
with the relations
\be
\chi_{1j} = \psi_j;\qquad \chi_{2j} = \sigma_3 \psi_j;\qquad \chi_{3j} = \sigma_3 \sigma_1 \psi^*_j;\qquad \chi_{4j} = \sigma_1 \psi^*_j,
\ee
and the orthonormality conditions read as
\be
\chi^\dagger_{\alpha j}\chi_{\beta k} =\delta_{\alpha\beta}\delta_{jk}.\label{ovortho}
\ee
Note that $(\chi_{1j},\chi_{2j})$ and $(\chi_{3j},\chi_{4j})$ are both chiral pairs and this is due to the fact that chiral symmetry is exactly preserved on the lattice within the overlap formalism~\cite{Narayanan:1994gw,Neuberger:1997fp}. The double degeneracy of the spectrum,
namely, $(\chi_{1j},\chi_{3j})$ and $(\chi_{2j},\chi_{4j})$, is due to the fermions being coupled to a real representation of the gauge group.
Presence of chiral modes of $V$ are associated with $\pm 1$ eigenvalues and they are also doubly degenerate:
\be
V\chi_{10}=\chi_{10};\qquad V\chi_{30}=\chi_{30};\qquad \sigma_3 \chi_{10}=-\chi_{10};\qquad \chi_{30}=\sigma_1\chi^*_{10};\qquad \sigma_3\chi_{30}=\chi_{30}
\ee
or
\be
V\chi_{20}=-\chi_{20};\qquad V\chi_{40}=-\chi_{40};\qquad \sigma_3 \chi_{20}=\chi_{20};\qquad \chi_{40}=\sigma_1\chi^*_{20};\qquad \sigma_3\chi_{40}=-\chi_{40}.
\label{zeroeig}
\ee
The doubly degenerate spectrum of $V$ shows that the overlap formalism for Dirac fermions in a real representation of the gauge group factorizes into two copies of Majorana fermions and the technical details are provided in~\apx{ovform}. The factorization of a Majorana fermion into a left and right handed Majorana-Weyl fermion has been discussed in~\cite{Huet:1996pw} and we repeat this discussion in the notation of this paper in~\apx{MW}. The zero modes that arise from the chiral modes are not a consequence of global gauge field topology. Yet, such background gauge field configurations have a Majorana fermion determinant of zero and one needs to insert one right-handed and one left-handed Majorana fermion to absorb this zero mode.  The role played by the presence of a zero mode is also discussed in~\apx{MW}.
The number of eigenvalues equal to $1$ and $-1$ mod($4$) are separately robust under perturbations and this separates the lattice gauge fields into two sectors. The details of the numerical computation for $SU(2)$ gauge fields with massless fermions in a real representation are presented in \scn{latform}. The main results of the paper are presented in \scn{mod2}, \scn{topcond} and \scn{nospont}.
We will numerically provide evidence for the following results in the continuum limit at fixed physics volume ($L\to\infty$ at fixed $\ell$)
for fermions in representations, $J=1,2,3,4$:
\begin{enumerate}
    \item Statistically speaking, all configurations with periodic boundary for gauge fields and fermions and all configurations with twisted boundary conditions for gauge fields and anti-periodic boundary conditions for fermions have zero modes. Therefore, in the context of \eqn{diffpart}, $Z_{++}$ and $Z_{--}$ are zero in the continuum limit. Numerical evidence is shown in \scn{mod2}.
    \item Statistically speaking, no configuration with periodic boundary for gauge fields and anti-periodic boundary conditions for fermions and all configurations with twisted boundary conditions for gauge fields and periodic boundary conditions for fermions have zero modes. Therefore, in the context of \eqn{diffpart}, $Z_{+-}$ and $Z_{-+}$ are non-zero. Numerical evidence is shown  in \scn{mod2}.
    \item Results for the  topological bilinear condensate at finite $\ell$ are presented in \scn{topcond} and we numerically show that $\langle M_i\rangle$; $i=1,2,3,4$, defined in \eqn{topcond} are finite and non-zero for all $\ell > 0$. In addition all of them approach the same value in the limit of $\ell\to\infty$ and the density per color degree of freedom seems to be independent of $J$, the representation of the Majorana fermion.
    \item  One could study spontaneous symmetry breaking by introducing a mass term and studying the massless limit. Such an approach requires one to first establish the small mass range and then study the behavior in that range. A better option from the numerical perspective is to study the spectral density which carries the same information. The spectral density of the eigenvalues of the Majorana fermion indicates the presence of a $\delta$-function at zero eigenvalue in sectors where there are no zero modes. This suggests the need to sum over partition functions to properly reproduce clustering property. Furthermore, there is no evidence for an additional final spectral density at zero eigenvalue. In addition,
    the ordered eigenvalues  given by
    \be
    \Lambda_j = \coth \frac{\phi_j}{2} \ge 0;\qquad j=1,2,\cdots;\qquad \Lambda_{j+1} > \Lambda_{j}
    \ee
    can be used to compute
    \be
    \lambda_{j+}(\ell) \ell = \lim_{L\to\infty} \frac{ \int [dU] e^{S^{z=1}_g(U)} (\Pf_a H_M ) (\Lambda_j L)}{\int [dU] e^{S^{z=1}_g(U)} (\Pf_a H_M)};\qquad \lambda_{j-}(\ell) \ell = \lim_{L\to\infty}\frac{ \int [dU] e^{S^{z=-1}_g(U)} (\Pf_p H_M) (\Lambda_j L)}{\int [dU] e^{S^{z=-1}_g(U)} (\Pf_p H_M)}
    \ee
    where the subscripts $p,a$ stand for periodic and anti-periodic boundary conditions for fermions.
We will numerically show that
    \be
    \lambda_{1\pm}(\ell) = \ell^{-1-\gamma_{m\pm}};\qquad 0 < \gamma_{m\pm} < 1.
    \ee
    This along with the behavior of the spectral density will provide strong evidence for the absence of spontaneous symmetry breaking of $Z_2$.
\end{enumerate}
All of the above will be shown to be valid for $J=1,2,3,4$ suggesting that the behavior of the fermion bilinear is not affected by the absence or presence of a conformal sector. These results will motivate us to study a theory of independent plaquettes that correspond to setting $z=0$ in the gauge action. We will find that both sectors (with and without zero modes) will survive the continuum limit and results discussed above can be obtained with a single partition function. The results so obtained will match the infinite volume results stated before for the topological condensate and the absence of spontaneous symmetry breaking.

\section{Overlap formalism with fermions in a real representation}\label{sec:overlap}

Let $U_\mu(x)$ denote a lattice link field in the real representation, namely,
\be
U_\mu^*(x) = U_\mu(x);\qquad U_\mu^t(x) V_\mu(x) = \mathbf{I}.
\ee
The parallel transporters are defined as
\be
(T_\mu \phi)(x) = U_\mu(x) \phi(x+\hat\mu)\quad\Rightarrow\quad (T^\dagger_\mu \phi)(x) = U^t_\mu(x-\hat\mu) \phi(x-\hat\mu);\qquad T_\mu^\dagger T_\mu= \mathbf{I};\qquad T_\mu^* = T_\mu.\label{paratrans}
\ee
The Wilson term is real and given by
\be
W = 2-m_w -\frac{1}{2}(T_1+T_2 + T_1^\dagger + T_2^\dagger);\qquad W^t = W;\qquad W^*=W,
\ee
and the two naive terms are also real and given by
\be
C_i = \frac{1}{2}(T_i - T_i^\dagger);\qquad C^t_i = -C_i;\qquad C_i^* = C_i;\qquad i=1,2.
\ee
The Hermitian Wilson-Dirac operator is
\be
H_w = \sigma_3( W + \sigma_1 C_1+\sigma_2 C_2);\label{Hwilson}
\ee
which can written as
\be
H_w
= \begin{pmatrix} W & C \cr C^\dagger & -W \end{pmatrix};\qquad C = C_1 - i C_2;\qquad C^t =-C.
\ee

\subsection{Spectral decomposition of $H_w$}\label{sec:specHw}
We start from \eqn{Hwilson} and observe
\be
\sigma_1 H_w \sigma_1 = -\sigma_3( W + \sigma_1 C_1-\sigma_2 C_2) = -H_w^*. \label{hwhwc}
\ee
Therefore,
\be
H_w \psi_i = \lambda_i \psi_i \quad\Rightarrow\quad H_w (\sigma_1 \psi^*) = - \lambda_i (\sigma_1 \psi^*),\label{Hwpair}
\ee
and we have fixed the phase of one eigenvector with respect to its pair. 
We will assume $\lambda_i > 0$.
We write
\be
H_w R
= R
\begin{pmatrix} \Lambda & 0 \cr 0 & -\Lambda \end{pmatrix};\qquad \Lambda_{ij} = \lambda_i \delta_{ij};\qquad R =\begin{pmatrix} A & B^* \cr B & A^* \end{pmatrix}\label{Hwspec}
\ee
Noting that
\be
R^* = \sigma_1 R \sigma_1 \label{ustaru}
\ee
follows from \eqn{hwhwc}, we see that
our choice of the unitary diagonalizing matrix, $R$, is special unitary.

\subsection{Spectral decomposition of $V$}\label{sec:specV}

The unitary operator
\be
V = \sigma_3 R \sigma_3 R^\dagger,
\ee
is a function of the gauge field and measures the rotation of $\sigma_3$ as a function of the gauge field.
Note that 
\be
V^\dagger = \sigma_3 V\sigma_3
\ee
and therefore
\be
V\psi_j = e^{i\phi_j}\psi_j \quad\Rightarrow\quad V\sigma_3\psi_j = e^{-i\phi_j}\sigma_3\psi_j\label{Veig}
\ee
which implies $\sigma_3\psi_j$ is an eigenvector of $V$ with eigenvalue $e^{-i\phi_j}$ and is orthonormal to $\psi_j$ as long as $\phi_j\ne 0,\pi$.
Using \eqn{ustaru}, the unitary operator, $V$, obeys
\be
V^* = \sigma_3 R^*\sigma_3 R^t = \sigma_1 V \sigma_1
\quad\Rightarrow V^\dagger = \sigma_1 V^t \sigma_1;\qquad V = \sigma_3\sigma_1 V^t \sigma_1\sigma_3
\quad\Rightarrow\quad \sigma_3 V\sigma_1 = -(\sigma_3 V \sigma_1)^t.\label{Vmajcond}
\ee
Taking the complex conjugate of \eqn{Veig} and using \eqn{ustaru}, we have
\be
\sigma_3 R^t \psi^*_j = e^{-i\phi_j}R^t\sigma_3 \psi^*_j \quad\Rightarrow\quad \sigma_1\sigma_3 R^\dagger (\sigma_1 \psi^*_j) =
e^{-i\phi_j} \sigma_1 R^\dagger\sigma_3 (\sigma_1 \psi^*_j)\quad\Rightarrow\quad \sigma_3 R^\dagger (\sigma_1 \psi^*_j) =
e^{-i\phi_j} R^\dagger\sigma_3 (\sigma_1 \psi^*_j)
\ee
implying that $\sigma_3\psi_j^*$ is an eigenvector with eigenvalue $e^{-i\phi_j}$ and this in turn implies that $\sigma_3\sigma_1\psi^*_j$ is an
 eigenvector with eigenvalue $e^{i\phi_j}$.
If the spectrum is not doubly degenerate, then
\be 
\psi_j = e^{i\theta}\sigma_3\sigma_1 \psi^*_j \quad\Rightarrow\quad \psi^*_j = e^{-i\theta}\sigma_3\sigma_1\psi_j = e^{-i\theta}\sigma_3\sigma_1 e^{i\theta}\sigma_3\sigma_1 \psi^*_j = -\psi^*_j \quad\Rightarrow\quad \psi^*_j=\psi_j=0.
\ee
Therefore, the spectrum is doubly degenerate even when the eigenvalue is zero.
Now we show that $\psi_j^\dagger \sigma_3\sigma_1 \psi_j^*=0$.
Assume this is not the case. Then we can set
\be
\psi_j^\dagger \sigma_3\sigma_1 \psi_j^* = a
\ee
We can perform a complex conjugate and a hermitian conjugate on the left-hand side resulting in
\be
\psi_j^t \sigma_3\sigma_1 \psi_j = a^*;\qquad \psi_j^t \sigma_1\sigma_3 \psi_j = a^* \quad\Rightarrow\quad a^* = -a^* \quad\Rightarrow\quad a=0.
\ee

The results can be written as 
\be
\chi_{1j} = \psi_j;\qquad \chi_{2j} = \sigma_3 \psi_j;\qquad \chi_{3j} = \sigma_3 \sigma_1 \psi^*_j;\qquad \chi_{4j} = \sigma_1 \psi^*_j,
\ee
with 
\be
V\chi_{1j} = e^{i\phi_j}\chi_{1j};\qquad
V\chi_{2j} = e^{-i\phi_j}\chi_{2j};\qquad
V\chi_{3j} = e^{i\phi_j}\chi_{3j};\qquad
V\chi_{4j} = e^{-i\phi_j}\chi_{4j}.
\ee
The orthonormality conditions read as
\be
\chi^\dagger_{\alpha j}\chi_{\beta k} =\delta_{\alpha\beta}\delta_{jk}.
\ee
The only freedom left in the choice of eigenvectors are
\be
\psi_j \to e^{i\alpha_j}\psi_j;\quad {\rm and}\quad (\chi_1,\chi_2) \leftrightarrow (\chi_3,\chi_4),\label{evfreedom}
\ee
where $\alpha_j$ is the remaining choice of phase.
If we have $\phi_0=0$ then
\be
V\chi_{10}=\chi_{10};\qquad V\chi_{30}=\chi_{30};\qquad \sigma_3 \chi_{10}=-\chi_{10};\qquad \chi_{30}=\sigma_1\chi^*_{10};\qquad \sigma_3\chi_{30}=\chi_{30}
\ee
and
if we have $\phi_0=\pi$ then
\be
V\chi_{20}=-\chi_{20};\qquad V\chi_{40}=-\chi_{40};\qquad \sigma_3 \chi_{20}=\chi_{20};\qquad \chi_{40}=\sigma_1\chi^*_{20};\qquad \sigma_3\chi_{40}=-\chi_{40}.
\ee
We can include the zero modes in the full orthonormality condition, namely, \eqn{ovortho}.
The eigenvalues of $V$ equal to $\pm 1$ are independently chirally paired. This is where the regulator plays a role. Let us assume that we work on a $L\times L$ lattice. With the fermion in the integer representation, let us define $N=(2J+1)L^2$.
The size of $V$ is $(2N\times 2N)$. If $L$ is even (odd) the size is (not) a multiple of $4$. Let us assume $L$ is even. If we have a multiple of $4$ eigenvalues of $V$ equal to $\pm 1$, we can assume that they can be lifted to values of $\phi\in (0,\pi)$ by a small perturbation of the gauge field. Therefore, we will call configurations with only two pairs of eigenvalues with one pair having $\phi=0$ and another having $\phi=\pi$ to correspond to a topological sector that is disconnected from configurations where all eigenvalues are $\phi\in (0,\pi)$. The pairing of $V=-1$ eigenvalues with $V=1$ eigenvalues is due to the rigid size of the matrix $V$. If $L$ was odd, then every configuration with either have one pair of $V=-1$ eigenvalues or one pair of $V=1$ eigenvalues.
We can write the spectral decomposition of $V$ as
\be
V = \sum_j \left[ e^{i\phi_j} \left( \chi_{1j}\chi^\dagger_{1j}
+ \chi_{3j}\chi^\dagger_{3j}\right)
+e^{-i\phi_j} \left( \chi_{2j}\chi^\dagger_{2j}
+ \chi_{4j}\chi^\dagger_{4j}\right)
\right] +  \left[ \left( \chi_{10}\chi^\dagger_{10}
+ \chi_{30}\chi^\dagger_{30}\right)
- \left( \chi_{20}\chi^\dagger_{20}
+ \chi_{40}\chi^\dagger_{40}\right)
\right].\label{Vspec}
\ee

\section{Details of the lattice formalism}\label{sec:latform}

This section is restricted to $SU(2)$ valued gauge fields since all our numerical simulations are for this choice.
Let us consider 2D lattice gauge theory on a periodic lattice of size $L^2$ for even values of $L$. Let $U^g_1(n_1,n_2)$ and $U^g_2(n_1,n_2)$
be $SU(2)$ values link variables in the positive $1$ and $2$ directions emanating from the site, $(n_1,n_2)$; $n_1,n_2\in \mathbf{Z}$. 
The single plaquette operator is
\be
P(n_1,n_2) = \left[ U^g_1(n_1,n_2) U^g_2(n_1+1,n_2) {U^g}_1^\dagger(n_1,n_2+1) {U^g}_2^\dagger(n_1,n_2)\right]
\ee
The Wilson gauge action can be written as~\cite{tHooft:1979rtg,GarciaPerez:1989gt}
 \be
 S(\beta,z) = \beta \sum_{n_1=0}^{L-1}\sum_{n_2=0}^{L-1} \Tr P(n_1,n_2) +(z-1)\beta P(L-1,L-1)\label{wilaction}
 \ee
where $z=1$ and $z=-1$ realizes un-twisted and twisted boundary conditions respectively on a periodic lattice and the link variables obey periodic boundary conditions in both cases. The case of $z=0$ is interesting since it corresponds to a theory with independent plaquettes as there is no global constraint arising from the last plaquette~\footnote{This is best seen by gauge fixing to 
\be
U_1^g(n_1,n_2) = 1;\quad 0\le n_1 < L-1;\qquad U_2^g(0,n_2) = 1 ;\quad0\le n_2 < L-1.
\ee
If we set $U^g_1(L-1,0)=W^g_1$ and $U^g_2(0,L-1)=W^g_2$, then all the other link variables are independently fixed by the $L^2-1$ plaquettes that does not include $P(L-1,L-1)$. The Wilson loop formed by winding around both directions of the torus is $W^g=W^g_1W^g_2{W_1^g}^\dagger {W_2^g}^\dagger$ and this is constrained by the plaquette, $P(L-1,L-1)$. Since $z=0$, the measure on $W^g_1$ and $W^g_2$ is just $d W^g_1\ dW^g_2$.
}
. We will set
\be
\beta=\frac{L^2}{\ell^2}
\ee
as mentioned in \scn{intro} and keep the dimensionless size, $\ell$, fixed as we take the continuum limit, $L\to\infty$. In addition, our data will show that results as a function of $\ell(2J+1)$ suggested by the analysis at large values of $J$ in~\cite{Kaushal:2023ezo} is relevant even at $J=1$.

Given a gauge field, $U^g$, we have  
\be U^g =  a_0 + ia_k\sigma_k = e^{i\theta_k \sigma_k}\quad\Rightarrow\quad a_0 = \cos\theta;\qquad a_k = \frac{\sin\theta}{\theta}\theta_k;\qquad \theta=\sqrt{\theta_k^2}.
\ee
The fermions couple to
\be
U=e^{2i\theta_k L_k};\qquad [L_j,L_k]=i\epsilon_{jkl}L_l;\qquad L_k^\dagger = L_k.
\ee
Given the standard form of the generators,
\be
T_3|J,M\rangle = M|J,M\rangle,\ \  T_\pm|J,M\rangle = \sqrt{J(J+1)-M^2\mp M}|J,M\pm 1\rangle,\ \ T_\pm = T_1\pm iT_2.
\ee
in the $|J,M\rangle$ basis for $M=-J,-J+1,\cdots, J-1,J$, basis.
we use~\cite{Narayanan:2023jmi}
\be
 L_a  = R T_a R^\dagger
\ee
where the unitary transformation $R$ is different from a unit matrix for integer representations
and
the non-zero elements  are 
\begin{eqnarray}
R_{M,M} & = &  \frac{1}{\sqrt{2}},\quad R_{-M,M}=\frac{i}{\sqrt{2}},\quad R_{M,-M}=\frac{(-1)^{M}}{\sqrt{2}},\cr
R_{-M,-M}& = &\frac{(-1)^{M-1}i}{\sqrt{2}} \ \ {\rm{for}} \quad M > 0,\qquad R_{0,0}=1.
\end{eqnarray}
The elements of the resulting $L_a$  are purely imaginary ($L_a=-L_a^t$) and $U$ is a real matrix that satisfies
\be
U^tU=\mathbf{I}
\ee
in the integer representations. We will consider fermions with periodic and anti-periodic boundary conditions and both these choices explicitly satisfy the reality property of the gauge fields seen by fermions that is necessary for the spectral properties derived in \scn{overlap}.

For the range of $\ell$ we will need to study the physics of interest to us, we found it optimal to generate gauge fields using just the gauge action in~\eqn{wilaction} and treat the fermionic determinant as an observable~\footnote{This method is inspired by the results obtained for the fermionic condensate in the Schwinger model using a similar method~\cite{Narayanan:1995sv}.}.
It is sufficient to use a standard heat-bath algorithm to generate a sequence of configurations. Since we want to approach small values of $\ell$, we need to ensure that the zero modes of the gauge fields are properly sampled.
We also need to ensure that the set of gauge fields that were generated were independent. To this end,  each configuration was obtained from a random cold start that was thermalized using one heat-bath step followed by one over-relazation step. We used~\cite{Kennedy:1985nu} for the heat-bath step and~\cite{Creutz:1987xi,Adler:1981sn,Adler:1987ce,Neuberger:1987gd} for the overrelaxation step. We ran the algorithm for enough pairs of heat-bath and overrelaxation steps to ensure that we are well into the thermalized regime. We could have continued taking measurements after thermalization but to ensure that the starting point was well sampled for the zero modes of gauge fields, we restarted after every measurement and ensured thermalization for each measurement. Since the main cost of the computation was in the evaluation of the fermionic eigenvalues, we could afford to add this minimal extra cost. We set all $U^g_k(n_1,n_2)=1$ except
\be
U_1^g(L-1,n_2)=W^g_1;\qquad U_2^g(n_1,L-1)=W^g_2;\qquad W^g_1W^g_2{W^g_1}^\dagger {W^g_2}^\dagger =1.
\ee
We can set $W^g_1$ to be diagonal with entries $e^{\pm i \alpha_1}$ with a gauge choice and $W^g_2W^g_1=W^g_1W^g_2$ will say that
$W^g_2$ is also diagonal with entries $e^{\pm i \alpha_2}$. The Haar measure is
\be
 d\alpha_1 d\alpha_2 \sin^2(\alpha_1) \sin^2(\alpha_2);\qquad \alpha_1,\alpha_2 \in [0,\pi].
 \ee
This was the procedure for generating the starting configuration. The partition functions $Z_{zf}$ with $(zf)=(++)$ and $(zf)=(+-)$ corresponds to setting $z=1$ in \eqn{wilaction} and imposing periodic or anti-periodic boundary conditions of fermions and $(zf)=(-+)$ and $(zf)=(--)$ corresponds to setting $z=-1$ in \eqn{wilaction} and imposing periodic or anti-periodic boundary conditions of fermions. We also ensured that we are in the correct continuum phase where Polyakov loops in both directions had averages consistent with zero.

\section{mod($2$) index on the lattice and the continuum}\label{sec:mod2}


\bef
  \centering
   \includegraphics[width=\linewidth]{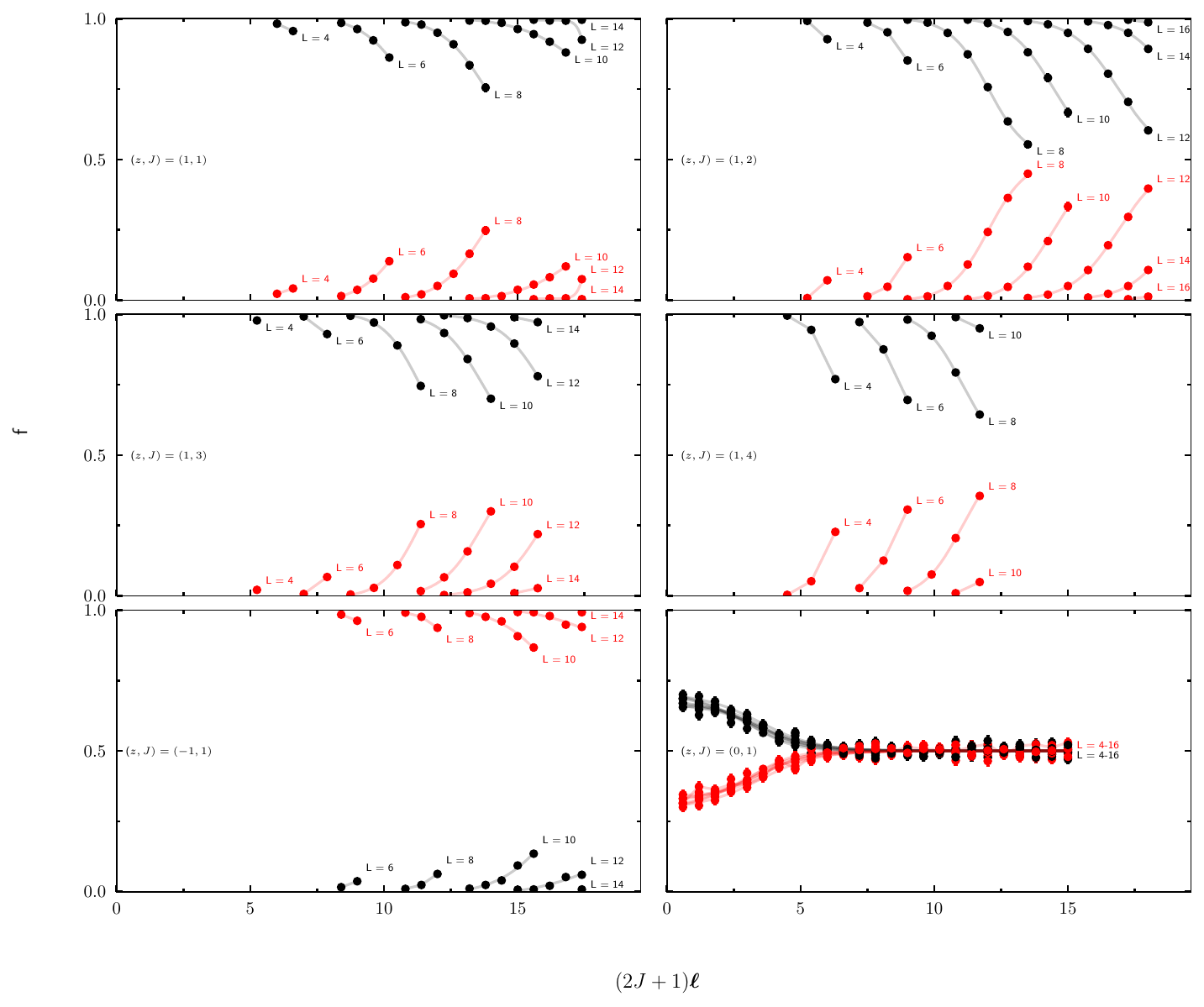}

  \caption{A plot of the fraction of zero modes as a function of $\ell$ at different values of $L$. The results for $(z,J)=(1,1),(1,2),(1,3),(1,4),(-1,1)$ ar shown in the different panels. The black and red points are for fermions with periodic and anti-periodic boundary conditions respectively.  } 
\eef{zeros}


Having generated gauge fields according to the measure solely given by the Wilson gauge action in~\eqn{wilaction}, we can define a quantity, $f$, as the fraction of the total number of configurations with zero modes for the massless overlap Dirac operator, $D(0)$, in \eqn{ovmassive}
which are configurations with eigenvalues of $V$ exactly equal to $-1$. This result will depend upon the choice of $z$, the representation, $J$, of the fermions and their boundary conditions. Five of the six panels in \fgn{zeros} shows the result for $(z,J)=(1,1),(1,2),(1,3),(1,4),(-1,1)$. We note that there are values of $\ell$ and $L$ where this fraction is not zero or one. Given a value of $L$, $\beta=\frac{L^2}{\ell^2}$ gets smaller as $\ell$ increases and this is an increase in the lattice spacing. The effect of this is to take $f$ away from from zero or one. When $z=1$, we note that $f$ moves away from one and zero for fermions with periodic and anti-periodic boundary conditions, respectively. This effect of finite lattice spacing seems to get stronger as $J$ increases but in all cases $f$ approaches one and zero in the continuum ($L\to\infty$ at a fixed $\ell$) for fermions with periodic and anti-periodic boundary conditions, respectively. When $(z,J)=(-1,1)$, we note that $f$ moves away from one and zero for fermions with anti-periodic and periodic boundary conditions, respectively. This could have been repeated for $J=2,3,4$. We note that $f$ approaches one and zero in the continuum ($L\to\infty$ at a fixed $\ell$) for fermions with anti-periodic and periodic boundary conditions, respectively. The behavior in the bottom-right panel is different. Here we have set $z=0$ in \eqn{wilaction} resulting in the presence of zero modes for both types of boundary conditions in the continuum limit when the gauge action is a product of independent plaquettes. Configurations with periodic boundary conditions for fermions are favored to have zero modes and configurations with anti-periodic boundary conditions for fermions are favored to have no zero modes as $\ell$ gets smaller. Since the measure associated with the Wilson loop that winds around both directions on the torus has no dependence on the gauge coupling $f\ne 0$ or $f\ne 1$ even as $\ell\to 0$.
In the limit of $\ell\to\infty$, the action with $z=0$ equally samples untwisted ($z=1$) and twisted ($z=-1$) gauge fields and half the configurations have zero modes and other half have no zero modes. There is a dependence of $f$ on $\ell$ when $z=0$ in contrast to $z=\pm 1$.

\section {A mod($2$) topological condensate in finite volume}\label{sec:topcond}

\bef
  \centering
  \includegraphics[width=\linewidth]{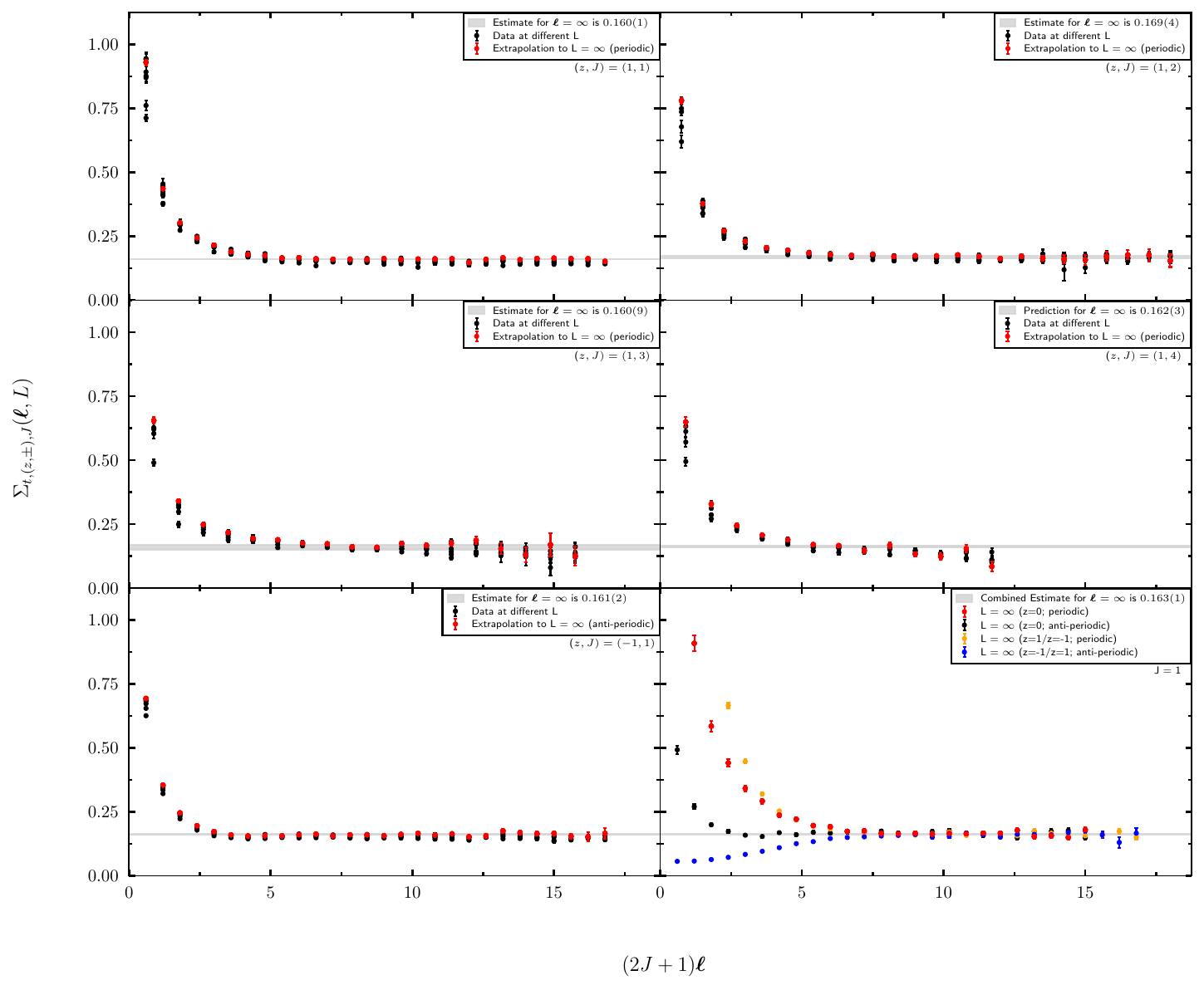}

  \caption{Extraction of $\Sigma_{t+}(\ell)$  from lattice data for Majorana fermions in the $J=1,2,3,4$ representations.} 
\eef{top}

Having generated gauge fields according to the measure solely given by the Wilson gauge action in~\eqn{wilaction}, we compute the spectrum of $V$ on each such configuration. This results in a list of $\phi_j\in (0,\pi)$ for each configuration along with each configuration being classified as one with or without zero modes. We can use \eqn{majdet} to compute the fermion determinant for a single massless Majorana fermion and hence measure
\be
 \frac{\int [dU^g] e^{S(\beta,z)} \prod_j \left[ \cos^2\frac{\phi_j}{2}\right]}{\int [dU^g] e^{S(\beta,z)}}\quad {\rm and}\quad
  \frac{\int [dU^g] e^{S(\beta,z)} \prod'_j \left[ \cos^2\frac{\phi_j}{2}\right]}{\int [dU^g] e^{S(\beta,z)}}\label{parfuns}
\ee
where the prime on the product implies that zero modes are not included when present. We will refer to these measurements as 
\be
{\bar Z}_{+-}(\ell,L);\quad {\bar Z}_{++}(\ell,L);\quad {\bar Z}_{--}(\ell,L);\quad {\bar Z}_{-+}(\ell,L)\quad{\rm and}\quad
{\bar Z}'_{+-}(\ell,L);\quad {\bar Z}'_{++}(\ell,L);\quad {\bar Z}'_{--}(\ell,L);\quad {\bar Z}'_{-+}(\ell,L)
\ee
for $z=1$ and anti-periodic boundary conditions for fermions; 
$z=1$ and periodic boundary conditions for fermions; 
$z=-1$ and anti-periodic boundary conditions for fermions; and
$z=-1$ and periodic boundary conditions for fermions respectively. 
These are related to the partition functions defined in \scn{intro} with a normalization that comes from the partition function of the pure gauge action.
As per the analysis in \scn{mod2} we have
\be
\lim_{L\to\infty} {\bar Z}_{++}(\ell,L) = \lim_{L\to\infty} {\bar Z}_{--}(\ell,L)
=\lim_{L\to\infty} {\bar Z}'_{+-}(\ell,L) = \lim_{L\to\infty} {\bar Z}'_{-+}(\ell,L) =0.
\ee
The quantities defined in \eqn{topcond} can be written as
\be
\langle M_1 \rangle = \frac{\bar Z'_{++}(\ell,L)}{\bar Z_{+-}(\ell,L)};\qquad
\langle M_2 \rangle = \frac{\bar Z'_{--}(\ell,L)}{\bar Z_{-+}(\ell,L)};\qquad
\langle M_3 \rangle = \frac{\bar Z'_{--}(\ell,L)}{\bar Z_{+-}(\ell,L)};\qquad
\langle M_4 \rangle = \frac{\bar Z'_{++}(\ell,L)}{\bar Z_{-+}(\ell,L)}.
\ee
With this in mind, we define topological condensates defined via \eqn{topcondpercfg} in \eqn{topcond} as
\be
\Sigma_{t,1,J}(\ell,L) =  \frac{1}{2m_w (2J+1)\ell L}\frac{ {\bar Z}'_{++}(\ell,L)}{\bar Z_{+-}(\ell,L)};\qquad
\Sigma_{t,-1,J}(\ell,L) =  \frac{1}{2m_w (2J+1)\ell L}\frac{ {\bar Z}'_{--}(\ell,L)}{\bar Z_{-+}(\ell,L)}\label{sigtop}
\ee
where the $\pm 1$ refer to the fixed value of $z$. We will primarily study these two quantities. In addition, we can also define
\be
\Sigma_{t,-,J}(\ell,L) =  \frac{1}{2m_w (2J+1)\ell L}\frac{ {\bar Z}'_{--}(\ell,L)}{\bar Z_{+-}(\ell,L)}\frac{\int [d U^g] e^{S(\beta, -1)}}{\int [dU^g]e^{S(\beta,1)}};\quad
\Sigma_{t,+,J}(\ell,L) =  \frac{1}{2m_w (2J+1)\ell L}\frac{ {\bar Z}'_{++}(\ell,L)}{\bar Z_{-+}(\ell,L)}\frac{\int [d U^g] e^{S(\beta, 1)}}{\int [dU^g]e^{S(\beta,-1)}}\label{sigtop1}
\ee
where the $\pm$ refers to the fixed value of the fermion boundary conditions. The extra factors of the pure gauge partition functions compensate for the factors in the denominators of \eqn{parfuns}.
All four of these quantities correspond to inserting a local fermion bilinear operator and the factors in front convert it into a density per color degree of freedon in the continuum and also takes the trivial wavefunction renomalization of the overlap-Dirac operator~\cite{Edwards:1998wx} into account.

Since these quantities are obtained as ratios of observables, we have used the jackknife method to estimate the biased average and the errors at a fixed $L$ and $\ell$. We have used at least three even values of $L$  at a fixed value of $\ell$ to extrapolate to $L\to\infty$
using a correction  of the form $\frac{1}{L^2}$. 
The results are shown in five panels of \fgn{top}. The black points in \fgn{top} show the right-hand side of \eqn{sigtop} for $\Sigma_{t,z,J}(\ell,L)$.  The red points are the extrapolated values at $L=\infty$. It is clear that the topological condensate is finite at any finite $\ell$ and it is also finite and non-zero as $\ell\to\infty$. We have sufficient number of points for $(z,J)=(1,1),(1,2),(1,3),(-1,1)$ where the condensate is independent of $\ell$ and we have used the results in that range to estimate a value for 
\be
\Sigma^\infty_{t,z,J} = \lim_{\ell\to\infty} \lim_{L\to\infty} \Sigma_{t,z,J}(\ell,L). 
\ee
With our normalization of the topological condensate being per color degree of freedom,   the value as $\ell\to\infty$ is independent of $z$ or $J$ and it is 
\be
\Sigma^\infty_{t,z,J} = 0.162(3).
\ee
Our numerical technique of treating the fermion determinant as an observable only works up to some values of $\ell$ since the fluctuations in the fermion determinant grows with $[(2J+1)\ell]^2$. We could not go to large enough values at $J=4$ to obtain an independent estimate of $\Sigma^\infty_{t,1,4}$ but the results seem to consistent with $0.162(3)$ obtained from the other values of $(z,J)$.

It might be more natural to consider mixed gauge boundary conditions and keep the fermion boundary condition fixed. These correspond to the quantities in \eqn{sigtop1} and the numerator and denominator are computed with different gauge actions ($z=1$ or $z=-1$).
The ratios of gauge actions with two different gauge boundary conditions can be computed analytically, namely,
 \be
 \lim_{L\to\infty} \frac{\int [d U^g] e^{S(\beta, 1)}}{\int [dU^g]e^{S(\beta,-1)}}  = \frac{\sum_{n=1}^\infty (-1)^{n-1} e^{- C_n \ell^2} }
 {\sum_{n=1}^\infty  e^{- C_n \ell^2}  };\qquad C_n = \frac{4n^2-1}{16}\label{partratio}
 \ee
 and we see that it approaches unity as $\ell\to\infty$. The results for $\Sigma_{t,\pm,1}(\ell,\infty)$ are shown in the bottom-right panel of \fgn{top}. We have only shown the results after extrapolation to $L\to\infty$ to avoid cluttering the plot. The behavior at finite $\ell$ are qualitatively different from the ones in the other panels but the result as $\ell\to\infty$ is numerically consistent with the ones in the other panels. The main contribution to the difference at finite $\ell$ arises from the $\ell$ behavior of \eqn{partratio}. 

The behavior with fixed fermion boundary conditions and mixed gauge boundary conditions in the bottom-right panel of \fgn{top} suggests that the case of $z=0$ is interesting. As per \eqn{parfuns}, we have
\be
{\bar Z}_{0-}(\ell,L);\quad {\bar Z}_{0+}(\ell,L);\quad{\rm and}\quad
{\bar Z}'_{0-}(\ell,L);\quad {\bar Z}'_{0+}(\ell,L)
\ee
and all of them are non-zero in the continuum limit.
To be consistent with \eqn{sigtop}, we define
\be
\Sigma_{t,0,J}(\ell,L) =  \frac{1}{2m_w (2J+1)\ell L}\frac{ {\bar Z}'_{0+}(\ell,L)}{\bar Z_{0+}(\ell,L)};\quad {\rm and}\quad
\Sigma_{t,0,J}(\ell,L) =  \frac{1}{2m_w (2J+1)\ell L}\frac{ {\bar Z}'_{0-}(\ell,L)}{\bar Z_{0-}(\ell,L)};
\label{sigtopz0}
\ee
with periodic and anti-periodic boundary conditions for fermions respectively.
Only configurations with zero modes  for fermions  contribute to the numerator and only configurations without zero modes for fermions contribute to the denominator.
The results using \eqn{sigtopz0} are also shown in the bottom-right panel of \fgn{top} for both boundary conditions of fermions only in the $L\to\infty$ limit.  Now the behavior at finite $\ell$ is affected by the dependence of $f$ on $\ell$ as shown in \fgn{zeros}. Since all four results in the bottom-right panel show the same quantitative behavior for large $\ell$, we were able to perform a combined asymptotic fit and the value is consistent with the ones from the other panels.

Using our results for $(z,J)=(1,1)$ we estimate the infinite volume topological condensate in units of the lightest mass~\footnote{We have used the value from Table 1 in~\cite{Narayanan:2023jmi} for $J=1$ and obtained a value for the lightest mass in our convention as $\sqrt{\frac{5.7 \times 8}{\pi}}=3.81$ where the factor of $8$ is needed to convert the coupling used in the light-cone quantization formalism in~\cite{Narayanan:2023jmi} to the one used in the Lagrangian formalism of this paper and the factor of $\pi$ arises since light-cone momenta were measured in units of $\pi$ in~\cite{Narayanan:2023jmi}.}  for $J=1$ to be
\be
3\frac{\Sigma_{t,+,1}(\infty)}{M_f} = 3\times\frac{0.162(3)}{3.81} = 0.128(2).
\ee
This roughly matches the value found in~\cite{Dempsey:2023fvm} once we note that our measurement of $\langle \psi_L\psi_R\rangle$ should be half the value for $\langle\psi^t\sigma_2 \psi\rangle$ computed in~\cite{Dempsey:2023fvm}.

\section{Absence of spontaneous symmetry breaking}\label{sec:nospont}

\bef
  \centering
  \includegraphics[width=\linewidth]{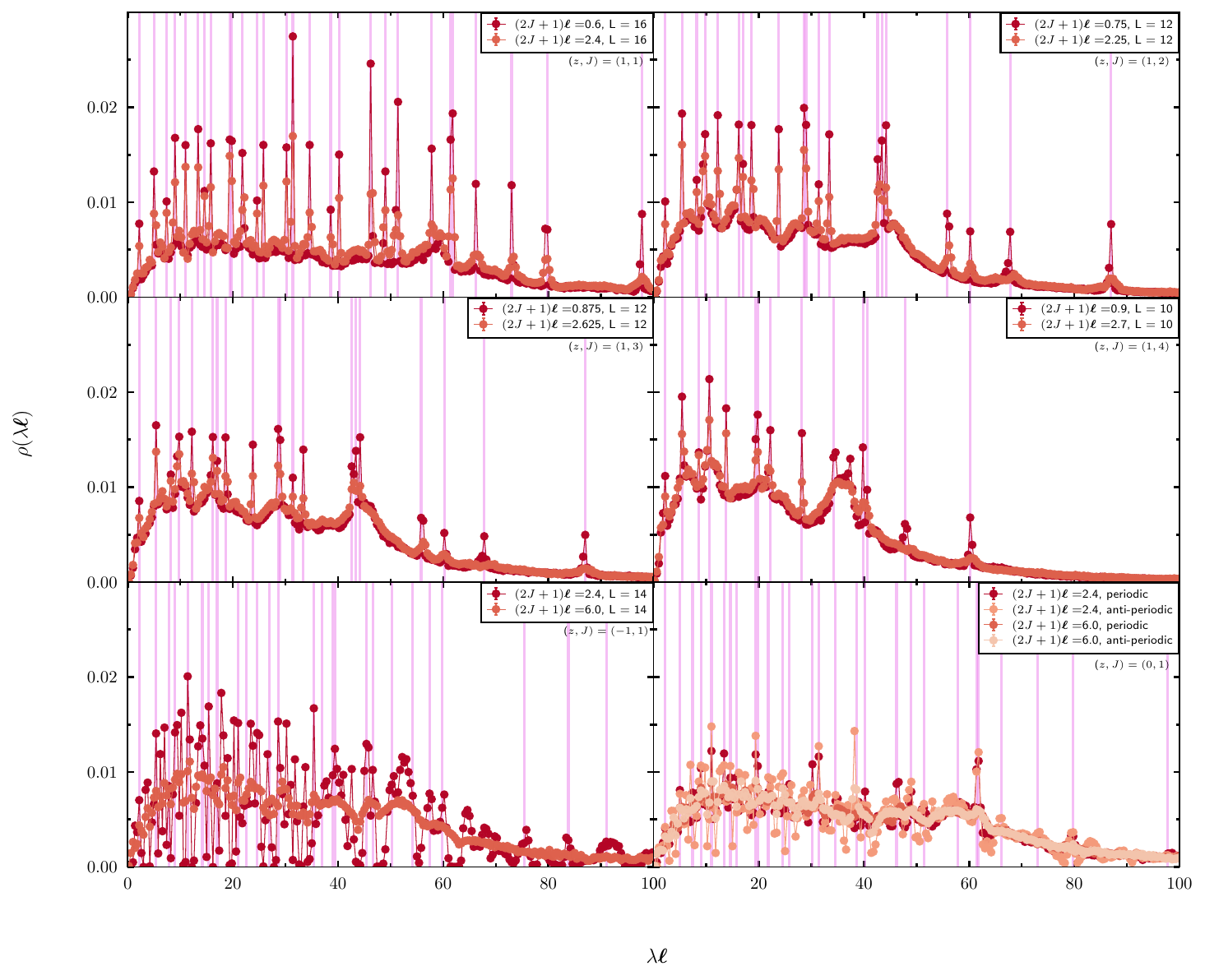}
  \caption{The behavior of the spectral density at small volumes for various choices of $(z,J)$.}
\eef{densitys}
\bef
  \centering
  \includegraphics[width=\linewidth]{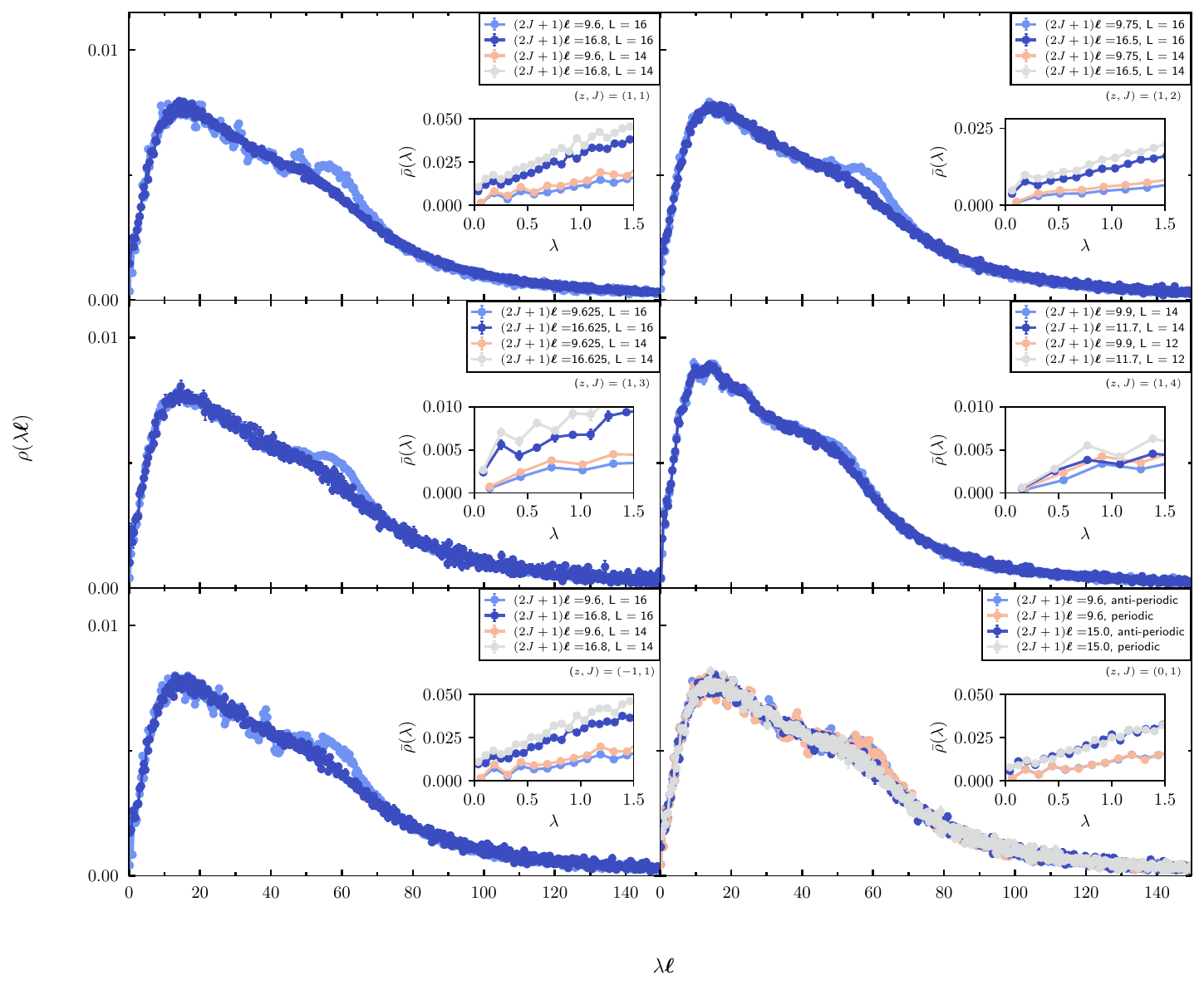}
  \caption{The behavior of the spectral density at small volumes for various choices of $(z,J)$. The inset shows the behavior close to zero.}
\eef{densityl}
\bef
  \centering
  \includegraphics[width=\linewidth]{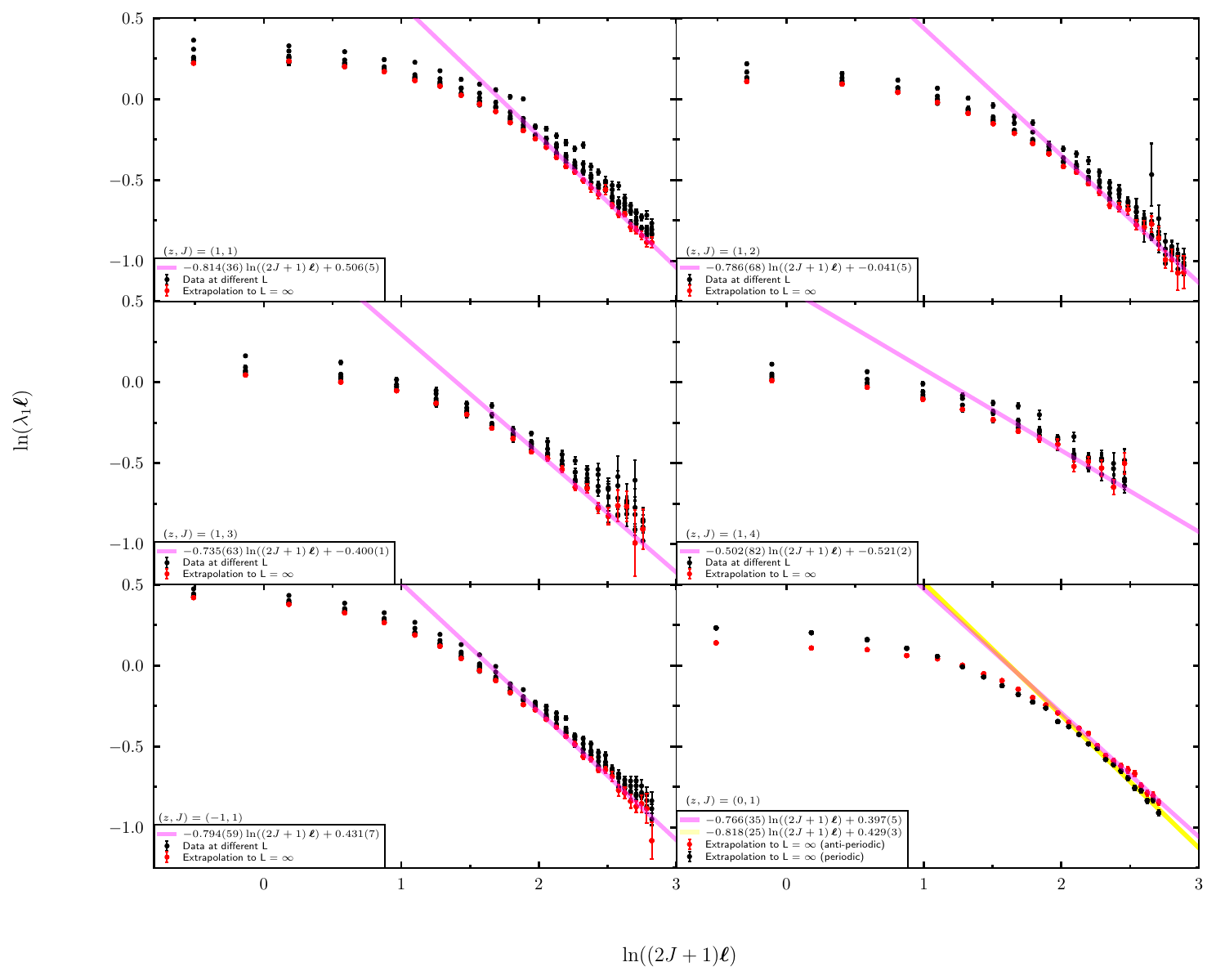}
  \caption{The behavior of the lowest eigenvalue, $\lambda_1(\ell)$, as function of $\ell$ for various choices of $(z,J)$.}
\eef{lowest}

We now consider the possibility that there can be a spontaneous breaking of the $Z_2$ symmetry in addition to the breaking from the presence of the zero modes discussed in~\scn{topcond}. For this purpose, we will focus on the spectrum of fermions with appropriate boundary conditions that have no zero modes in the continuum limit. 
Given a fixed value of $(z,L,J)$, we can compute the entire fermion spectrum on each gauge field configuration. With this information, we can obtain a distribution, $\rho(\lambda\ell)$, at a fixed value of $\ell$ with the normalization
\be
\int_0^\infty \rho(\lambda\ell) d[\lambda\ell] = \frac{1}{2};\qquad \rho(-\lambda\ell)=\rho(\lambda\ell).
\ee
This should show a transition from small values of $\ell$ where it should show deviations from free field behavior to a smooth behavior at large values of $\ell$. We found that the distribution approaches a limit as $L$ is increased and we only show the results at a fixed value of $L$ that is large enough to explain the behavior. Let us first look at the panels corresponding to $(z,J)=(1,1),(1,2),(1,3),(1,4),(-1,1)$ in \fgn{densitys}.
The vertical lines correspond to the locations in the free field limit. We observe sharp peaks matching the location of the free field values at the smallest value of $\ell$ in each of the five panels and these peaks start to spread out as $\ell$ is increased. Now, let us move to the results shown for $(z,J)=(1,1),(1,2),(1,3),(1,4),(-1,1)$ in \fgn{densityl}. The smallest value of $\ell$ shown in each of the five panels exhibit peaks. This value of $\ell$ is roughly in the range where the topological condensate shown in \fgn{top} is about to reach a plateau in the large $\ell$ limit. The two higher values of $\ell$ shown in the five panels do not show evidence for peaks and the distribution seems to have settled on a large $\ell$ limit. This behavior is not typical since we are looking at $\rho(\lambda\ell)$ as a function of $(\lambda\ell)$. Since
\be
\rho(\lambda\ell) d(\lambda\ell) = \bar\rho(\lambda) d\lambda;\qquad \bar\rho(\lambda) = \ell\rho(\lambda\ell) \ee
we see that $\bar\rho(\lambda)$ viewed as a function of $\lambda$ is expected to have a $\delta$-function in its distribution at $\lambda=0$. This behavior is not consistent with a spontaneous symmetry breaking since the fermion bilinear would diverge as $\frac{1}{m}$ which signals an emergence of zero modes in the $\ell\to\infty$ limit that is prohibited at finite $\ell$. This could signal the space of the torus being split into two with one sector capable of producing a zero mode while the other sector does not. The insets in each of the panels focus on the behavior of $\bar\rho(\lambda)$ close to $\lambda=0$. Consistent with a sharpening of the peak approaches zero, we see that the density gets larger with $\ell$ in all six panels at a fixed $\lambda$. But we see a finite lattice spacing effect that has a consistent tendency of reducing the density with decreasing lattice spacing. Since we are not in a position to perform a systematic analysis of the finite lattice spacing effects (this will need a significantly increased statistics and results at larger values of $L$), we come to the rough conclusion that there is probably no other contribution at $\lambda=0$ other that a $\delta$-function in the $\ell\to\infty$ limit. 

Since a non-zero value of $\rho(0)$ cannot be convincingly ruled out,  we consider the observable defined in \eqn{smallev},
\be
\langle\Lambda_1\rangle(L,\ell) =  \frac{\langle \cot\frac{\phi_1}{2} \rangle_{+-}}{{\bar Z}_{+-}};\qquad \langle\Lambda_1\rangle(L,\ell) =  \frac{\langle \cot\frac{\phi_1}{2} \rangle_{-+}}{{\bar Z}_{-+}}. 
\ee
This quantity is also a ratio of observables and we use the jackknife method to estimate the errors at finite $L$ and $\ell$. The limit to infinite $L$ defined in \eqn{smallev} is obtained by fitting the data with a leading correction of the form $\frac{1}{L^2}$. The results for $(z,J)=(1,1),(1,2),(1,3),(1,4),(-1,1)$ are shown in the five panels of \fgn{lowest}. The flattening of the data for $(2J+1)\ell < 2$ is consistent with $\rho(\lambda\ell)$ for small $\ell$ and shoes that we have reached the small $\ell$ limit where the eigenvalues behavior is close to that of free field. The data for $(2J+1)\ell> 4$ suggests that we have reached the asymptotic limit and the behavior is consistent with
\be
\lambda_1 \sim \ell^{-1-\gamma_m(z,J)},
\ee
with
\be
\gamma_m(1,1)=0.814(36);\quad \gamma_m(-1,1)=0.794(59);\quad \gamma_m(1,2)=0.786(68);\quad \gamma_m(1,3)=0.735(63);\quad \gamma_m(1,4)=0.502(82).
\ee
The values for $\gamma_m(\pm 1,1)$ are consistent with each other. The higher statistics at $(z,J)=(1,1)$ shows that a value of $1$ is ruled out by $4.5$ standard deviations. A value of $1$ at $(z,J)=(1,2)$ is ruled out by $3.3$ standard deviations even with an error that is larger by a factor of $2$ compared to $(z,J)=(1,1)$. Finally a value of $1$ is clearly ruled out for $(z,J)=(1,3),(1,4)$. In addition we also see a decreasing trend in $\gamma_m$ as $J$ is increased. We should conclude our analysis of \fgn{densitys} and \fgn{densityl} by pointing that the results for $(z,J)=(0,1)$ shown in the bottom-right panel of both the figures behave in a manner similar to the other five panels and result in values for $\gamma_m$ consistent with the other two values for $J=1$.

 The combined analysis of the spectrum density and scaling behavior of the lowest eigenvalue has ruled out a condensate from a finite density of the eigenvalue spectrum at $\lambda=0$ and we see no evidence for spontaneous symmetry breaking. An alternate approach used in~\cite{Bergner:2024ttq} is to compute the fermion bilinear as a function of the fermion mass. Wilson fermions used in~\cite{Bergner:2024ttq} do not exhibit a mod($2$) index and as such it is not possible to separate explicit symmetry breaking from spontaneous symmetry breaking. Furthermore, the symmetry associated with the spectrum of $H_w$ derived in \scn{specHw} further complicates matters regarding additive mass renormalization with Wilson fermions. Taking into account that $\gamma_m=0.817(39)$ for $J=1$, any non-analytic behavior in mass will occur close to the massless limit. As such, we opt not to compare our results obtained bere with the ones obtained in~\cite{Bergner:2024ttq}.

\section{Discussion and conclusions}

We have studied two dimensional $SU(2)$ gauge theories coupled to a single Majorana fermion in a real representation labelled by an integer value of $J$. We have used a lattice formalism that realizes fermions using the overlap formalism. The Wilson gauge action is given in \eqn{wilaction} and
has a parameter $z$ in addition to the lattice gauge coupling, $\beta$. A value of $z=1$ realizes gauge fields with periodic boundary conditions and $z=-1$ realizes gauge fields with a non-trivial twisted boundary conditions. 

The gauge field dependent unitary operator $V$ defined in \eqn{Vop} plays a crucial role in our analysis and appears in the formula for the overlap-Dirac operator in \eqn{ovmassive}. All eigenvalues of $V$ are doubly degenerate as a consequence of the real representation of the Dirac fermion. This facilitates a factorization of a single Dirac fermion into two Majorana fermions as explained in \scn{ovform}. Furthermore, exact chirality on the lattice implies a pairing of eigenvalues of $V$, namely, $e^{i\phi_j}$; $ \phi_j \in (0,\pi)$ with its chiral partner, $e^{-i\phi_j}$. The eigenvectors are chiral if $\phi_j=0$ or $\phi_j=\pi$
and each of these eigenvalues, if present, are doubly degenerate and chirally paired. This results in the possibility of a mod($2$) index first proposed in~\cite{Huet:1996pw} and later analyzed in~\cite{Cherman:2019hbq}. The factorization of a Majorana fermion into a pair of Majorana fermions is explained in \scn{MW} where the mod($2$) index is shown in the context of the operator $V$.

Since exact zero modes of the massless Dirac operator will severely affect a standard Hybrid Montecarlo algorithm with dynamical fermions, we decided to generate gauge fields using just the Wilson action and treat the fermion determinant as part of an observable. This will work as long as the physical volume of the two-dimensional continuum torus is not large and we are able to extract infinite volume results using our restricted finite volume analysis. 

Our first observation is that an action with $z=1$ results in all continuum gauge fields with periodic (anti-periodic) boundary conditions for fermions to have a non-zero (zero) mod($2$) index and this is consistent with the continuum analysis in~\cite{Cherman:2019hbq}. The role of fermion boundary conditions is reversed when $z=-1$. If we were to define a theory with a partition function that is not zero, we could then address whether the $Z_2$ chiral symmetry present in such a theory is spontaneously broken. 

The lattice Hamiltonian analysis in~\cite{Dempsey:2023fvm} shows the presence of a bilinear fermion condensate. In order to understand this result in the Lagrangian formalism, we postulate a topological bilinear condensate as  the expectation value of the fermion bilinear from exact zero modes using mixed fermion boundary conditions -- with $z=1$ ($z=-1$) we compute ratios of expectation value of fermion determinant with periodic (anti-periodic) boundary conditions and expectation value of fermion determinant with anti-periodic (periodic) boundary conditions. It is assumed that fermion determinants are computed without the zero modes and this accounts for an insertion of a fermion bilinear as discussed in \scn{MW}.
Mixed fermion boundary conditions were easy in our numerical simulations since expectation values appearing in the ratio are with respect to the same gauge action. We are also able to study expectation values with mixed gauge boundary conditions since we have an analytic formula for the partition function with different gauge boundary conditions. We show in \scn{topcond} that the topological bilinear condensate as defined is non-zero in any finite volume and the value obtained at infinite volume matches the value obtained in~\cite{Dempsey:2023fvm} for $J=1$. In addition, this topological condensate exists for $J=2,3,4$ and does not differentiate between the low energy spectrum of theory -- $J=1,2$ have a mass gap while $J=3,4$ have an infinite massless sector~\cite{Delmastro:2021otj,Narayanan:2023jmi}. The four different ways of defining the topological condensate using mixed boundary conditions show different behavior at finite $\ell$ but all of them reach the same limiting value as $\ell\to\infty$. Finally, the topological condensate per color degree of freedom comes out to be independent of $J$.

 The results obtained in this paper are counter-intuitive. Gauge field boundary conditions or fermion boundary conditions are usually not expected to affect the final physics results -- they are usually used to suppress finite volume effects or control the effects of trivial fermion zero modes. We have shown that it drastically affects the partition function in the continuum limit to the extent that it is not possible to show evidence for a non-zero fermion bilinear condensate by fixing the boundary conditions for gauge and fermion fields. One way out is to sum over boundary conditions of gauge fields noting that it is not the same as summing over topological sectors although the presence of the mod($2$) index suggests an analogy. 
 We have the option of summing over fermion boundary conditions or gauge field boundary conditions. Since it might be more natural to sum over gauge field boundary conditions with fixed fermion boundary conditions, we may consider the standard gauge action with the extra parameter $z$ in \eqn{wilaction} not necessarily set to $\pm 1$. In a two dimensional non-abelian gauge action, the non-trivial constraint on the Wilson loop obtained by winding on both directions of the torus can be localized to the flux at a point on the torus. The translation to a Wilson gauge action is simple -- one of the $L^2$ plaquettes on the lattice is responsible for the global Wilson loop that winds around both directions of the torus. If we set the coupling of that plaquette to zero ($z=0$ in \eqn{wilaction}), then we have an action with independent plaquettes and a measure on the global Wilson loop that is independent of the gauge coupling. How can changing the contribution of one plaquette to the lattice action matter in the continuum limit? Contrary to the standard actions with $z=\pm 1$, where all gauge field configurations either produce zero modes or not depending on the boundary conditions, action with $z=0$ results in a fraction $f$ of the configurations to have zero modes or not depending on the boundary conditions and this fraction changes from a value away from $\frac{1}{2}$ at small volumes to $\frac{1}{2}$ at large volumes. This difference is seen when one compares the bottom-right panel ($z=0$) of \fgn{zeros} to the other five panels ($z=\pm 1$). With $z=0$, we can compute the topological condensate without mixing boundary conditions. We find that such a condensate is non-zero at all sizes of the continuum torus for both choices of boundary conditions of fermions implying that the symmetry is broken explicitly.

 To further understand the physics of zero modes, we studied
the spectral distribution in the sector without zero modes in \scn{nospont}. This does not require summing over partition functions. The spectral distribution viewed as $\rho(\lambda\ell)$ where $\lambda$ is the Dirac eigenvalue and $\ell$ is the size of the torus shows good evidence of approaching a limit as $\ell\to \infty$ (see \fgn{densityl}). This implies a contribution from the $\delta$-function at zero eigenvalue in the infinite volume limit for $\bar\rho(\lambda)$ suggesting an emergence of zero modes even when it is prohibited at any finite volume and not evidence for spontaneous symmetry breaking. To confirm the absence of spontaneous symmetry breaking, we also studied the scaling of the lowest eigenvalue as a function of the size of torus and found no evidence for a spontaneous symmetry breaking.

Our analysis suggests the need to address the issue of clustering in correlation functions. Consider, for example, the partition function with $z=1$ and anti-periodic boundary conditions for fermions. Based on our analysis, we expect the two-point function of the fermion bilinear to not approach zero as the separation goes to infinity~\footnote{This is not easy to see in a finite volume analysis but Nikhil Karthik saw some evidence for this in a earlier thought process in this project.}. If indeed this is the case, we have to conclude that the one-point function of the fermion bilinear is not zero and we would say that this comes from choosing periodic boundary conditions for fermions or setting $z=-1$ in the action for gauge fields. Analogous to the need to sum over gauge field topologies to restore clustering the two dimensional Schwinger model, we would have to sum over all boundary conditions for fermions or gauge fields. An alternate approach would be to consider a gauge action with independent plaquettes ($z=0$ in \eqn{wilaction}) and this theory will have some gauge field configurations with zero modes and other with no zero modes implying a separation into two sectors labeled by the mod($2$) index akin to gauge field topology. There will be a fermion bilinear condensate at all finite volume and we would say that the $Z_2$ symmetry is explicitly broken by the mod($2$) index. Clustering of two point functions will be restored when the sum over the two mod($2$) sectors and taken into account.

\acknowledgments
The authors thank Nikhil Karthik for some preliminary work on this project and for extensive discussions on the role played by boundary conditions of gauge fields. The authors thank Sruthi Narayanan for several physics discussions and a careful reading of the manuscript.
The authors also thank Simon Hands and Erich Poppitz for useful discussions. R.N. acknowledges partial support by the NSF under grant number
PHY-2310479.  This work used Expanse at SDSC through allocation
PHY240084 from the Advanced Cyberinfrastructure Coordination
Ecosystem: Services \& Support (ACCESS) program, which is supported
by National Science Foundation grants \#2138259, \#2138286, \#2138307,
\#2137603, and \#2138296.
\appendix
\section{Overlap-Dirac formalism for Majorana fermions}\label{sec:ovform}
The overlap formalism for a Dirac fermion in a real representation of the gauge group starts with the many-body Hamiltonians,
\be
{\cal H}_+ = \begin{pmatrix} a^\dagger & b^\dagger \end{pmatrix} H_w
\begin{pmatrix} a \cr b \end{pmatrix};\qquad
{\cal H}_- = -\begin{pmatrix} a^\dagger & b^\dagger \end{pmatrix}
\sigma_3
\begin{pmatrix} a \cr b \end{pmatrix} \label{ovHpm}
\ee
where the Hermitian Wilson-Dirac operator is given in \eqn{Hwilson} and $a$ and $b$ obey canonical anti-commutation relations.
Let $|0\pm\rangle$ be the normalized ground states of ${\cal H}_\pm$.
The generating functionals for the chiral fermion pair that make up a massless Dirac fermion are given by
\be
Z_L(\bar\eta_L,\eta_L) = {}_L\langle 0-|e^{\bar\eta_L a_L} e^{b_L^\dagger \eta_L} |0+\rangle_L;\qquad 
Z_R(\bar\eta_R,\eta_R) = {}_R\langle 0+|e^{\bar\eta_R b_R} e^{a_R^\dagger \eta_R} |0-\rangle_R.
\ee
Note that the order in which the exponential factors do not matter since they commute with each other. Furthermore, we have only included operators that propagate.
The details that result in formulas for $Z_L$ and $Z_R$ is similar to the one found in~\cite{Karthik:2016ppr} and we sketch the steps.
Let us use \eqn{Hwspec} and define
\be
c = A^\dagger a + B^\dagger b;\qquad d = B^t a + A^t b\quad\Rightarrow\quad c^\dagger = a^\dagger A + b^\dagger B;\qquad d^\dagger = a^\dagger B^* + b^\dagger A^*,
\ee
and note that $c,d$ obey anti-commutation relations.
Then,
\be
{\cal H}_+ = \sum_i \lambda_i (c^\dagger_i c_i - d^\dagger_i d_i);\quad {\cal H}_-=\sum_i (a_i^\dagger a_i -b^\dagger_i b_i);\quad\Rightarrow\quad
c_i |0+\rangle=0;\quad d^\dagger_i |0+\rangle=0;\quad a_i| 0-\rangle=0; \quad b_i^\dagger| 0-\rangle=0.
\ee
Let us form two copies of the above with $L,R$ being the subscripts. 
To evaluate $Z_L$ we write
\be
\bar\eta_L a_L + b_L^\dagger \eta_L = Q_+-Q_-;\qquad Q_+=\bar\eta_L (A^\dagger)^{-1} c_L + d_L^\dagger (A^*)^{-1}\eta_L;\qquad Q_-=\bar\eta_L (A^\dagger)^{-1}B^\dagger b_L + a_L^\dagger B^* (A^*)^{-1}\eta_L
\ee
Then
\be
[Q_+,Q_-] = -2\bar\eta_L G^\dagger \eta_L;\qquad
G= BA^{-1}.
\ee
and
similarly,
\be
Z_R(\bar\eta_R,\eta_R) =  e^{\bar\eta_R G \eta_R}{}_R\langle 0+|0-\rangle_R.
\ee
The mass term couples the left and right section and is realized as
\be
e^{m\left(a_R^\dagger a_L - b^\dagger_L b_R\right) } = \left[\int [d\bar\xi_L][d\xi_R]\ e^{-\bar\xi_L\xi_R +\sqrt{m}\left(\bar\xi_L a_L +a^\dagger_R \xi_R\right)}\right]  \left[\int [d\bar\xi_R][d\xi_L]\ e^{\bar\xi_R\xi_L +\sqrt{m}\left(\bar\xi_R b_R +b^\dagger_L \xi_L\right)}\right]  .
\ee 
This enables us to insert the factors in each of the chiral sectors and then perform the integrals over $\bar\xi_L,\bar\xi_R,\xi_L,\xi_R$. The  generating functional for a massive Dirac fermion is given by
\be
Z_V(\bar\eta_L,\eta_L,\bar\eta_R,\eta_R;m) =
 \det\left[m^2+(1-m^2)AA^\dagger\right] \ e^{\bar\eta {\begin{pmatrix} m & G^{-1} \cr -(G^\dagger)^{-1} & m \end{pmatrix}}^{-1}\eta};
\qquad
\bar\eta=\begin{pmatrix}-\bar\eta_L & \bar\eta_R\end{pmatrix};\quad
\eta=\begin{pmatrix}\eta_R \cr \eta_L\end{pmatrix}.
\ee

We now proceed to write the above results solely in terms of $V$ and $m$. To that end, the massive Dirac operator  is
\be
D(m) = \frac{1+m}{2} + \frac{1-m}{2} V.\label{ovmassive}
\ee
and one can show that
\be
\det D(m) = \det [m^2+(1-m^2)AA^\dagger];\qquad G^{-1}_V(m) = \begin{pmatrix} m & G^{-1} \cr -(G^\dagger)^{-1} & m \end{pmatrix}= \frac{(1+m)+(1-m)V}{1-V}.
\ee

Using \eqn{Vspec}, the spectral decomposition of the massive Dirac operator in \eqn{ovmassive} and the propagator are
\bea
D(m) 
&=& \sum_j \left[ \frac{(1+m)+(1-m)e^{i\phi_j}}{2} \left( \chi_{1j}\chi^\dagger_{1j}
+ \chi_{3j}\chi^\dagger_{3j}\right)
+\frac{(1+m)+(1-m)e^{-i\phi_j}}{2} \left( \chi_{2j}\chi^\dagger_{2j}
+ \chi_{4j}\chi^\dagger_{4j}\right)
\right] \cr
&& +  \left[ \left( \chi_{10}\chi^\dagger_{10}
+ \chi_{30}\chi^\dagger_{30}\right)
+m \left( \chi_{20}\chi^\dagger_{20}
+ \chi_{40}\chi^\dagger_{40}\right)
\right];\cr
G_V(m) 
&=& \sum_j \left[ \frac{1} {m+i\cot\frac{\phi_j}{2}}\left( \chi_{1j}\chi^\dagger_{1j}
+ \chi_{3j}\chi^\dagger_{3j}\right)
+\frac{1}{m-i\cot\frac{\phi_j}{2}} \left( \chi_{2j}\chi^\dagger_{2j}
+ \chi_{4j}\chi^\dagger_{4j}\right)
\right] \cr
&& +  \left[ 
\frac{1}{m} \left( \chi_{20}\chi^\dagger_{20}
+ \chi_{40}\chi^\dagger_{40}\right)
\right].
\eea
The second line in each of the expressions are present only if there are zero modes and we also notice the two-fold degeneracy of the spectrum that follows from \eqn{Vspec}.

To realize a single Majorana fermion, consider 
the spectral decomposition for $H_M(m) = \sigma_3 D(m)\sigma_1$, namely,
\bea
H_M(m) &=&  \sum_{j\ne 0}  \left[ \frac{(1+m)+(1-m)e^{i\phi_j}}{2} \left( \chi_{2j}\chi^t_{4j}
- \chi_{4j}\chi^t_{2j}\right)
+\frac{(1+m)+(1-m)e^{-i\phi_j} }{2}\left( \chi_{3j}\chi^t_{1j}
- \chi_{1j}\chi^t_{3j}\right)
\right]\cr
&&  +  m\left( \chi_{20}\chi^t_{40}
- \chi_{40}\chi^t_{20}\right)
+\left( \chi_{30}\chi^t_{10}
- \chi_{10}\chi^t_{30}\right).
\label{HMdecomp}
\eea
and it is explicit that $H_M(m)$ is an anti-symmetric matrix. Next we note that
\be
\det D(m) = \det H_M(m) = \int [d\bar\xi][d\xi] e^{\bar\xi H_M(m) \xi}.
\ee
If we define 
\be
\bar\xi = \frac{1}{\sqrt{2}}(\psi+\omega);\qquad \xi=\frac{1}{\sqrt{2}}(\psi-\omega)
\ee
the antisymmetric property of $H_M(m)$ results in
\be
\bar\xi H_M(m) \xi = \frac{1}{2}\psi H_M(m) \psi -\frac{1}{2}\omega H_M(m) \omega 
\ee
and a single Majorana fermion is realized as
\be
\int [d\psi] e^{\frac{1}{2}\psi H_M(m) \psi} = \Pf\left[\frac{1}{2} H_M(m)\right].
\ee
If we decompose
\be
\psi = \sum_{\alpha j}\psi_{\alpha j}\chi^*_{\alpha j}
\ee
then
\be
\frac{1}{2}\psi H_M(m) \psi = \sum_{j\ne 0} \left( \frac{ (1+m)+(1-m)e^{i\phi_j}}{2} \psi_{2j}\psi_{4j} 
+ \frac{(1+m)+(1-m)e^{-i\phi_j}}{2} \psi_{3j}\psi_{1j}\right)+  m\psi_{20}\psi_{40} 
+  \psi_{30}\psi_{10}.
\ee
If we define the Grassmann integral as $
\int \prod_{j} d\psi_{1j}d\psi_{2j}d\psi_{3j}d\psi_{4j}
$, then
\be
\Pf (H_M(m)) = m \prod_{j\ne 0} \left(\left[\cos^2\frac{\phi_j}{2}+m^2\sin^2\frac{\phi_j}{2}\right]\right) 
\label{majdet}
\ee
Note that the integral as defined is invariant under \eqn{evfreedom} and we can conclude that the Pfaffian is positive for a positive mass.
In a similar manner, we define $G_M(m) = \sigma_3 G_V(m) \sigma_1$ and obtain
\be
G_M(m) =  \sum_{j\ne 0} \frac{1} {m+i\cot\frac{\phi_j}{2}} \left[  \left( \chi_{2j}\chi^t_{4j}
- \chi_{4j}\chi^t_{2j}\right)
+\frac{1} {m-i\cot\frac{\phi_j}{2}}\left( \chi_{3j}\chi^t_{1j}
- \chi_{1j}\chi^t_{3j}\right)
\right]
  +  \frac{1}{m}\left( \chi_{20}\chi^t_{40}
- \chi_{40}\chi^t_{20}\right).
\label{GMdecomp}
\ee
which is also anti-symmetric.
Then we have
\be
\bar\eta G_V(m) \eta = \bar\eta \sigma_3 G_M(m) \sigma_1 \eta
\ee
and if we write
\be
\bar\eta \sigma_3 = \frac{1}{\sqrt{2}}(\phi +  \rho);\qquad \sigma_1\eta = \frac{1}{\sqrt{2}}(\phi-\rho)
\ee
then
\be
\bar\eta G_V(m) \eta = \frac{1}{2}\phi G_M(m) \phi -\frac{1}{2} \rho G_M(m) \rho,
\ee
showing the factorization into two Majorana fermions.
 
If we decompose
\be
\phi = \sum_{\alpha j}\phi_{\alpha j}\chi^*_{\alpha j}
\ee
then
\be
\frac{1}{2}\phi G_M(m) \phi = \sum_{j\ne 0}  \left(\frac{1}{m+i\cot\frac{\phi_j}{2}} \phi_{2j}\phi_{4j} 
+ \frac{1}{m-i\cot\frac{\phi_j}{2}} \phi_{3j}\phi_{1j}\right)
+ \frac{1}{m} \phi_{20}\psi_{40}.
\ee
The generating functional for a Majorana fermion is
\be
Z_M(\phi;m) = \Pf\left[\frac{1}{2}H_M(m)\right] e^{\frac{1}{2}\phi G_M(m) \phi}.
\ee

A Majorana bilinear condensate would be an average over the measure of the quantity
\be
\Sigma(m) = \sum_j \left[\frac{\partial}{\partial \phi_{4j}}\frac{\partial}{\partial \phi_{2j}} + 
\frac{\partial}{\partial \phi_{1j}}\frac{\partial}{\partial \phi_{3j}}\right] Z_m(\phi;m)\Bigg|_{\phi=0}
= \Sigma_t(m) + \sigma_s(m) 
\ee
where the topological condensate 
\be
\Sigma_t(m) = \prod_{j\ne 0} \left(\left[\cos^2\frac{\phi_j}{2}+m^2\sin^2\frac{\phi_j}{2}\right]\right) \label{topcondpercfg}
\ee
is only present when there are zero modes and 
\be
\Sigma_s(m) = 
\Pf(H_M(m)) \left[ \sum_{j\ne 0} \frac{2m}{m^2+\cot^2\frac{\phi_j}{2}} \right]\label{sponcondpercfg}
\ee
is the condensate due to spontaneous symmetry breaking when there are no zero modes.
The topological condensate is present on a finite torus since $\Sigma_t(0)$ is non-zero.
On the other hand, $\Sigma_s(0)=0$ and we need to study this quantity in the limit $\ell\to\infty$ followed by $m\to 0$.
Following the standard Banks-Casher argument~\cite{Banks:1979yr}, if the average over the gauge measure
\be
\lambda_j \ell = \lim_{L\to\infty} \left\langle \Lambda_j\right\rangle L; \qquad \Lambda_j=\cot\frac{\phi_j}{2}\label{smallev}
\ee
behaves as
\be
 \lambda_j = \Sigma z_j \ell^{-2} \label{evasym}
 \ee
 as $\ell\to\infty$ and $z_j$ are constants determined by chiral random matrix theory~\cite{Verbaarschot:2000dy}, then we would conclude that $\Sigma$ is the value of the condensate from spontaneous symmetry breaking.

\section{Majorana-Weyl fermions and the mod($2$) index}\label{sec:MW}

It is useful to start with \eqn{ovHpm} and show the factorization into Majorana-Weyl fermions. This will also help us understand how zero modes appear in spite of the fact that the spectrum of $H_w$ has paired positive and negative eigenvalues. In that sense the mechanism is different from zero modes arising from global gauge field topology~\cite{Narayanan:1993ss}.

The factorization of the many-body Hamiltonian into two identical parts can be shown by defining
\be
a= \frac{\xi-i\eta}{\sqrt{2}};\qquad b=\frac{\xi^\dagger-i\eta^\dagger}{\sqrt{2}} \quad\Rightarrow\quad \xi=\frac{a+b^\dagger}{\sqrt{2}};\qquad \eta=\frac{b^\dagger-a}{\sqrt{2}i}.
\ee
The fermion operators, $\xi$ and $\eta$, also obey canonical anti-commutation relations. Using the properties that $W=W^t$ and $C=-C^t$, we see that
\bea
{\cal H}_+ &=& \frac{1}{2}\begin{pmatrix} \xi^\dagger & \xi \end{pmatrix}
H_w
\begin{pmatrix} \xi \cr \xi^\dagger \end{pmatrix} + \frac{1}{2}\begin{pmatrix} \eta^\dagger & \eta \end{pmatrix}
H_w
\begin{pmatrix} \eta \cr \eta^\dagger\end{pmatrix}\cr
{\cal H}_- &=&  -\frac{1}{2}\begin{pmatrix} \xi^\dagger & \xi \end{pmatrix}
\sigma_3
\begin{pmatrix} \xi \cr \xi^\dagger \end{pmatrix} - \frac{1}{2}\begin{pmatrix} \eta^\dagger & \eta \end{pmatrix}
\sigma_3
\begin{pmatrix} \eta \cr \eta^\dagger\end{pmatrix},\label{Majfac}
\eea
This shows the factorization into two identical many-body Hamiltonians and we have (dropping the $L$ or $R$ subscript)
\be
|0\pm\rangle = |0\pm\rangle_\xi \times |0\pm\rangle_\eta;\qquad
\langle 0-|0+\rangle = {}_\xi\langle 0-|0+\rangle_\xi\  {}_\eta\langle 0-|0+\rangle_\eta= \left[{}_\xi\langle 0-|0+\rangle_\xi\right]^2,\label{chiralfac}
\ee
with each factor being associated with a single Majorana-Weyl fermion. 
Using the spectral decomposition in \eqn{Hwspec}, one half of the many-body Hamiltonian in \eqn{Majfac} can be written as
\be
{\cal H}_+ = -\frac{1}{2}\begin{pmatrix} \xi^\dagger & \xi \end{pmatrix}
\begin{pmatrix} A & B^* \cr B & A^* \end{pmatrix}
\sigma_3
\begin{pmatrix} A^\dagger & B^\dagger \cr B^t & A^t \end{pmatrix}
\begin{pmatrix} \xi \cr \xi^\dagger \end{pmatrix} = \begin{pmatrix} \xi^\dagger & \xi \end{pmatrix}
\begin{pmatrix}  AA^\dagger - \frac{1}{2}& AB^\dagger \cr -A^*B^t & \frac{1}{2}-A^*A^t\end{pmatrix}
\begin{pmatrix} \xi \cr \xi^\dagger\end{pmatrix};\qquad 
{\cal H_-} = \frac{1}{2} (\xi^\dagger\xi - \xi\xi^\dagger).
\ee

Referring to \eqn{Veig}, we write
\be
\psi_j = \frac{1}{\sqrt{2}}\begin{pmatrix} u_j \cr d_j\end{pmatrix}
\ee
as a normalized eigenvector with $\phi_j\ne 0,\pi$. Note that neither $u_j$ nor $d_j$ is a zero vector since
$\sigma_3\phi_j$ is orthogonal to $\phi_j$.
The orthogonality of $\psi_j$ and $\sigma_3\psi_j$ implies
\be
u_j^\dagger u_j = d_j^\dagger d_j=1.\label{udnorm}
\ee
The orthogonality of $\psi_j$ with $\sigma_1\psi_j^*$ and $\sigma_3\sigma_1\psi_j^*$ implies
\be
u_j^td_j = d_j^tu_j=0.\label{udortho}
\ee

Using \eqn{Hwspec}, \eqn{Vop} and \eqn{Veig} we have
\be
 B^\dagger d_j = i\tan\frac{\phi_j}{2} A^\dagger u_j;\quad
B^t u_j = i\tan\frac{\phi_j}{2} A^t d_j.
\ee
Using the unitarity of $R$ in \eqn{Hwspec} we arrive at
\bea
AA^\dagger u_j = \cos^2\frac{\phi_j}{2} u_j; & BB^\dagger u^*_j = \sin^2\frac{\phi_j}{2} u^*_j;& AB^\dagger d_j =\frac{i\sin\phi_j}{2}u_j;\cr
AA^\dagger d^*_j = \cos^2\frac{\phi_j}{2} d^*_j; & BB^\dagger d_j = \sin^2\frac{\phi_j}{2} d_j; & AB^\dagger u^*_j = -\frac{i\sin\phi_j}{2}d^*_j.\label{evrels}
\eea
If we have a $\phi=\pi$, it will correspond to a zero eigenvalue of $AA^\dagger$ and this will not be doubly degenerate as discussed after \eqn{zeroeig}.
Since the dimensions of $A$ and $B$ are $N=(2J+1)L^2$ and is even ($L$ is even) there will also be an eigenvalue with $\phi=0$ which corresponds to a unit eigenvalue of $AA^\dagger$. In this case, we will label the eigenvector with $\phi=\pi$ as $u_1$ and the eigenvector with $\phi=0$ as $d_1^*$ and label the remaining eigenvectors in increasing values of $\cos^2\frac{\phi_j}{2}$ with $j=2,\cdots,\frac{N}{2}$.
We will write the eigenspace of $AA^\dagger$ as
\be
X = \begin{pmatrix}u_1 & d^*_1 & u_2 & d^*_2 & \cdots u_{\frac{N}{2}}& d^*_{\frac{N}{2}} \end{pmatrix};\qquad 
AA^\dagger X = X\frac{\mathbf{I} + C}{2} ;\label{Xmat}
\ee
where $C$ is a diagonal matrix of size $N\times N$ given by
\bea
 C_{ij}&=& c_j \delta_{ij};\qquad c_{2j-1}=c_{2j}=\cos\phi_j;\qquad j=1,\cdots,\frac{N}{2};\cr
{\rm and}\quad   C_{ij}&=& c_j \delta_{ij};\qquad c_1=-1;\ c_2=1;\ c_{2j-1}=c_{2j}=\cos\phi_j;\qquad j=2,\cdots,\frac{N}{2}.
\eea
where there are no zero modes and when there is one zero mode respectively.
 Since we cannot reach the gauge field background that has a zero and unit of eigenvalue of $AA^\dagger$ from one that does not have the pair by a continuous deformation, the choice of order for the two sets can be different.

Let us also define the unitary matrix
\be
Y = X^*\Sigma;\qquad \Sigma^2=1;\quad \Sigma^t=\Sigma^\dagger=\Sigma;\qquad
\Sigma_{2j-1,2j}=\Sigma_{2j,2j-1}=1;\qquad j=1,\cdots,\frac{N}{2}.
\ee
Explicitly,
\be
Y=\begin{pmatrix}d_1 & u^*_1 & d_2 & u^*_2 & \cdots d_{\frac{N}{2}}& u^*_{\frac{N}{2}} \end{pmatrix}.
\ee
We can write the equations in \eqn{evrels} as
\be
A^*A^t X^* = X^* \frac{\mathbf{I}+C}{2};\qquad
AB^\dagger Y = X \frac{iS}{2}  ; \qquad
A^*B^t X = -X^*\frac{iS}{2} \Sigma
\ee
where $S$ is a diagonal matrix of size $N\times N$ given by
\bea
S_{jk} &=& s_j \delta_{jk};\qquad s_{2j-1}=-s_{2j}=\sin\phi_j;\qquad j=1,\cdots,\frac{N}{2}\cr
{\rm and}\quad S_{jk} &=& s_j \delta_{jk};\qquad s_1=s_2=0;\ s_{2j-1}=-s_{2j}=\sin\phi_j;\qquad j=2,\cdots,\frac{N}{2}
\eea
when $AA^\dagger$ has no zero eigenvalues and when it has one zero eigenvalue respectively.

Let us perform  a change of operators by
\be
\xi = X\omega = \omega X^t\quad\Rightarrow\quad \xi^\dagger = X^*\omega^\dagger = \omega^\dagger X^\dagger,
\ee
and we note that $\omega,\omega^\dagger$ obey canonical anti-commutation relations.
Then 
\be
 {\cal H}_- = \frac{1}{2}(\omega^\dagger\omega - \omega\omega^\dagger) ;\qquad 
{\cal H}_+ = \frac{1}{2}\begin{pmatrix} \omega^\dagger & \omega \end{pmatrix}
\begin{pmatrix}  C & iS\Sigma \cr iS\Sigma & -C\end{pmatrix}
\begin{pmatrix} \omega \cr \omega^\dagger\end{pmatrix}.
\ee
An explicit computation shows that the non-zero elements of $S\Sigma$ are
\be
(S\Sigma)_{2j-1,2j} = s_j;\qquad (S\Sigma)_{2j,2j-1}=-s_j;
\ee
for $ j=1,\cdots,\frac{N}{2}$, when $AA^\dagger$ has no zero or unit eigenvalue and
for $ j=2,\cdots,\frac{N}{2}$,
when $AA^\dagger$ has a zero and unit eigenvalue. Therefore,
\bea
{\cal H}_+ &=& \begin{cases}
{\cal H}_0 +\sum_{j=2}^{\frac{N}{2}}\left[ -\cos\phi_j + {\cal H}_{j}\right] & AA^\dagger{\rm\ has\ a\ unit\ and\ zero\ eigenvalue}\cr
\sum_{j=1}^{\frac{N}{2}}\left[ -\cos\phi_j + {\cal H}_{j}\right] & AA^\dagger{\rm\ has\ no\ unit\ or\ zero\ eigenvalue}\end{cases} \cr
{\cal H}_j &=& \cos\phi_j\left(
\omega^\dagger_{2j-1} \omega_{2j-1} +\omega^\dagger_{2j} \omega_{2j}\right) + i\sin\phi_j\left(
 \omega_{2j-1}^\dagger\omega^\dagger_{2j} 
+\omega_{2j-1}\omega_{2j} \right);\cr
{\cal H}_0&=&\frac{1}{2}-\omega_1^\dagger\omega_1.
\eea
The ground state of ${\cal H}_-$ is the trivial vacuum $|0\rangle$ that is annihilated by $\omega_j$.
The ground state of ${\cal H}_+$ can be written in a factorized form for each $j$. Let us define

The space acted upon by ${\cal H}_j$ with $j\ne 0$ is spanned by
\be
|1\rangle=|0\rangle,\quad |2\rangle=\omega_{2j-1}^\dagger |0\rangle;\qquad |3\rangle = \omega_{2j}^\dagger |0\rangle;\qquad |4\rangle=\omega_{2j-1}^\dagger \omega_{2j}^\dagger |0\rangle.
\ee
The action of ${\cal H}_j$ on these four states results in
\be
{\cal H}_j |1\rangle = i\sin\phi_j |4\rangle;\qquad {\cal H}_j |2\rangle = \cos\phi_j |2\rangle;\qquad {\cal H}_j|3\rangle = \cos\phi_j|3\rangle;\qquad
{\cal H}_j|4\rangle =-i\sin\phi_j|1\rangle +2\cos\phi_j|4\rangle.
\ee
It follows that
\bea
{\cal H}_j \left[ -i\sin\frac{\phi_j}{2}|1\rangle+\cos\frac{\phi_j}{2}|4\rangle\right] &=& (\cos\phi_j+1)\left[ -i\sin\frac{\phi_j}{2}|1\rangle+\cos\frac{\phi_j}{2}|4\rangle\right]; \cr
{\cal H}_j \left[ \cos\frac{\phi_j}{2}|1\rangle-i\sin\frac{\phi_j}{2}|4\rangle\right] &=& (\cos\phi_j-1)\left[ \cos\frac{\phi_j}{2}|1\rangle-i\sin\frac{\phi_j}{2}|4\rangle\right] .
\eea
Since $1+\cos\phi_j > \cos\phi_j > \cos\phi_j -1$ for all $\phi_j\in (0,\pi)$, we conclude that the ground state of ${\cal H}_+$ is unique when $AA^\dagger$ has no zero or unit eigenvalue and is given by
\be
|0+\rangle = \prod_{j=1}^{\frac{N}{2}} \left[ \cos\frac{\phi_j}{2} -i\sin\frac{\phi_j}{2} \omega_{2j-1}^\dagger \omega_{2j}^\dagger\right]|0\rangle.
\ee
The fermion determinant for a single Majorana-Weyl fermion in the absence of zero and unit eigenvalues of $AA^\dagger$ is
\be
\langle 0|0+\rangle = \prod_{j=1}^{\frac{N}{2}} \cos\frac{\phi_j}{2}.
\ee
Since ${\cal H}_+$ does not depend on $\omega_2$ or $\omega_2^\dagger$ when $AA^\dagger$ has a zero and unit eigenvalue, there are two degenerate ground states, namely
\bea
|0+\rangle_1 &=& \omega_1^\dagger\prod_{j=2}^{\frac{N}{2}} \left[ \cos\frac{\phi_j}{2} -i\sin\frac{\phi_j}{2} \omega_{2j-1}^\dagger \omega_{2j}^\dagger\right]|0\rangle;\cr
|0+\rangle_2 &=& \omega_1^\dagger\omega_2^\dagger\prod_{j=2}^{\frac{N}{2}} \left[ \cos\frac{\phi_j}{2} -i\sin\frac{\phi_j}{2} \omega_{2j-1}^\dagger \omega_{2j}^\dagger\right]|0\rangle.
\eea
This results in
\be
\langle 0|0+\rangle_1=\langle 0|0+\rangle_2=0;\qquad \langle 0|\omega_1|0+\rangle_1=\langle 0|\omega_2\omega_1|0+\rangle_2= \prod_{j=2}^{\frac{N}{2}} \cos\frac{\phi_j}{2} \equiv \langle 0|-+\rangle'_1\label{degenexp}
\ee
Note that $\omega_1^\dagger$ has to be inserted to obtain a non-zero expectation value and this is the operator associated with the zero mode.
The operator $\omega_2^\dagger$ is associated with the unit eigenvalue of $AA^\dagger$ and it is an {\sl ultra-violet} mode. Let us insert lattice spacing to convert from lattice operators to continuum operators. To this end, note that a mass term that couples a left and right Majorana fermion is
\be
M\sum_x \xi_L^t \eta_L = \frac{M}{a} \int d^2x \frac{\xi_L^t}{\sqrt{a}} \frac{\eta_L}{\sqrt{a}}.
\ee
Noting that $|0+\rangle_2$ has one extra fermion operator compared to $|0+\rangle_1$ we see that
a conversion from lattice operators to continuum operators will say that $|0+\rangle_2$ is suppressed by a factor of $\sqrt{a}$ compared to $|0+\rangle_1$ and we need to only consider $|0+\rangle_1$ as the relevant ground state.

\bibliography{biblio}
\end{document}